\documentclass{article}
\pdfoutput=1
\usepackage{jheppub}
\usepackage{hyperref}
\usepackage{graphicx}
\usepackage{datetime}
\usepackage{float}

\usepackage{amsmath}
\usepackage{amssymb}
\usepackage{amsthm}
\usepackage{mathrsfs}
\usepackage{slashed}
\usepackage{tabularx}
\usepackage{color}
\usepackage{comment}
\usepackage{multirow}
\usepackage{booktabs,graphicx}

\begin{document}

\author{Finn Larsen,} \author{Yangwenxiao Zeng}
\title{Black Hole Spectroscopy and AdS$_2$ Holography}
\affiliation{Department of Physics and Leinweber Center for Theoretical Physics, University of Michigan,\\450 Church
Street, Ann Arbor, MI 48109-1120, USA.}
\emailAdd{larsenf@umich.edu}\emailAdd{zengywx@umich.edu}

\abstract{
We compute the spectrum of extremal nonBPS black holes in four dimensions by studying supergravity on their AdS$_2\times S^2$ near horizon geometry.
We find that the spectrum exhibits significant simplifications even though supersymmetry is completely broken. We interpret our results in the framework of nAdS$_2$/nCFT$_1$ correspondence and by comparing with dimensional reduction from AdS$_3$/CFT$_2$ duality. As an additional test we compute quantum corrections to extremal black hole entropy on the nonBPS branch and recover results previously determined using very different methods.}


\maketitle


\section{Introduction and Summary}
An important step towards a detailed understanding of quantum black holes is the determination of their spectrum \cite{Maldacena:1996ix}. However, with the exception of BPS black holes, it has generally proven quite difficult to compute the black hole spectrum precisely. In this paper we find the spectrum of extremal nonrotating black holes on the nonBPS branch of ${\cal N}=8$ and  ${\cal N}=4$ supergravity.

The black holes we consider are solutions to  theories with extended supersymmetry and have AdS$_2\times S^2$ near horizon geometry, just like BPS black holes; but they are supported by fluxes that are inconsistent with supersymmetry. In this situation it is not expected that the spectrum is organized by supersymmetry and our explicit computations confirm this generic expectation. However, we find that nonetheless the {\it black hole spectrum exhibits significant simplifications} that are {\it reminiscent of the familiar ones that are due to supersymmetry}. This finding does not conform with textbook BPS-ology but we  will explain how it fits nicely with other expectations.

The spectrum of the black holes we consider is described by the quantum numbers of the $SL(2)\times SU(2)$ isometries of AdS$_2\times S^2$, i.e. the conformal weight $h$ and the partial wave number $j$. The conformal weight is equivalent to the mass $m$ of the perturbations in units of the AdS$_2$ radius $\ell$ through
\begin{equation}
\label{eqn:hform}
h = \frac{1}{2}+ \sqrt{\frac{1}{4}+ m^2\ell^2}~,
\end{equation}
for scalar fields. For BPS black holes the supersymmetry algebra guarantees that the supergravity mass spectrum corresponds to conformal weights $h$ that are integers for bosons and half-integers for fermions. For nonBPS black holes the masses of fluctuations in supergravity are not constrained a priori but our explicit computations establish that, in fact, the values of $m^2$ for scalar fields are all such that the conformal weights \eqref{eqn:hform} are integers. This is part of our claim that the spectrum is reminiscent of supersymmetry. In particular, the result suggests that the supergravity spectrum on the nonBPS branch is protected against quantum corrections and, if so, it should offer detailed guidance towards construction of the UV complete string theory describing extreme nonBPS black holes, despite the absence of supersymmetry.

The technical aspects of our explicit computations follow the strategy that is very well known from similar problems addressed in the past, such as spherical reduction of type IIB supergravity in ten dimensions on AdS$_5\times S^5$ \cite{Kim:1985ez,Duff:1986hr}. Accordingly, we first find the equations of motion of 4D supergravity and then linearize them around our AdS$_2\times S^2$ background solution. We then expand all fluctuating fields in their partial wave components and impose gauge conditions. It is no surprise that the 2D equations that result from these steps are messy, but fortunately they are sufficiently block diagonal that they can be disentangled and solved, despite the absence of supersymmetry. The final mass matrices therefore straightforwardly give eigenvalues for the masses of each partial wave that we can insert in \eqref{eqn:hform} and so identify the conformal weights in AdS$_2$.

The only subtlety that is special to two dimensions is the spin of the fields \cite{Camporesi:1994ga,Sen:2011ba,Larsen:2014bqa}. In AdS$_2$ we can generally represent vectors and tensors as scalar fields and similarly recast gravitinos as Majorana-Weyl fermions. However, the dualization of fields with with spin require special considerations for {\it harmonic modes} because those are generated by gauge symmetries that are ``large" in the sense that they are non-normalizable on AdS$_2$. Therefore, such transformations are not true symmetries, they generate field configurations that are physical and interpreted as excitations that are localized on the boundary. They can be identified with the modes that are described by a Schwarzian action (and its generalizations) in the Jackiw-Teitelboim 
model (and its relatives) \cite{Maldacena:2016upp,Engelsoy:2016xyb,Jensen:2016pah,Fu:2016vas,Moitra:2018jqs,Sachdev:2019bjn}. We refer to these modes as {\it boundary modes} following the terminology previously used in the context of logarithmic corrections to black hole entropy in four dimensions. Thus the spectrum of extremal black holes on the nonBPS branch is characterized by
\begin{itemize}
\item
{\it Bulk modes} that, from the AdS$_2$ point of view, are organized in infinite towers of Kaluza-Klein modes (partial waves).
\item
{\it Boundary modes} that, from the AdS$_2$ point of view, are field configurations that are physical even though they can be represented as ``pure gauge" locally. These modes are closely related to {\it harmonic modes}.
\end{itemize}
Our result for the quantum numbers of supergravity on the nonBPS branch of AdS$_2\times S^2$ are reported in table \ref{summary:h}. As a test of this spectrum we have computed the quantum contributions due to these modes by explicitly summing over all physical states. We find agreement with logarithmic corrections to the black hole entropy previously found using local methods \cite{Castro:2018hsc}. This gives great confidence in the black hole spectrum we find.

We have already mentioned that on the nonBPS branch all fields in AdS$_2$ have integral conformal weight $h$ and table \ref{summary:h} shows that we mean this quite literally: the {\it conformal weight is integral even for fermions}. This assignment is unusual but not inconsistent because the familiar relation between spin and statistics does not apply in two dimensions, at least in its standard form. Indeed, we will confirm our finding that fermions have integral weight on the nonBPS branch by recovering this assignment in settings where the AdS$_2$ geometry descends from an AdS$_3$ factor.

The standard simplification due to supersymmetry is that, when certain conditions are satisfied, the spectrum is organized into short multiplets
that enjoy some protection against quantum corrections. However, there is also a less frequently exploited simplification that is due to broken supersymmetry. On the BPS branch both simplifications are relevant but on the nonBPS branch it is only the latter one that applies. It can be interpreted as a global supersymmetry that is implemented directly on the black hole spectrum. We discuss this symmetry in detail in section \ref{section:global symmetry}.

Before getting to details of our computations we must carefully consider the {\it meaning} of the spectrum of quadratic fluctuations around AdS$_2\times S^2$. Indeed, several well-known results prompt the question of whether such a spectrum makes any sense at all. For example, \footnote{There are closely related results for the near horizon Kerr geometry and our discussion below should apply to that case as well \cite{Amsel:2009ev,Dias:2009ex}.}
\begin{itemize}
\item Finite energy excitations in AdS$_2$ are incompatible with asymptotically AdS$_2$ boundary conditions: they elicit strong gravitational backreaction that modifies the asymptotic structure of spacetime \cite{Maldacena:1998uz}. Therefore, quadratic fluctuations are not intrinsic to AdS$_2$.
\item 
In constructions where AdS$_2$ arises from AdS$_3$ through reduction along a null direction it was argued that the excitations with the lowest energy depend on the compact null coordinate but not on the AdS$_2$ that is retained by the compactification \cite{Balasubramanian:2009bg}. Therefore, the perturbations varying over AdS$_2$ that we consider do not dominate in the infrared limit.
\end{itemize}
In view of such results it is, for example, not obvious that the AdS$_2$ conformal weight $h$ is a useful quantum number in AdS$_2$ quantum gravity. However,
the recent development of nAdS$_2$/nCFT$_1$ correspondence \cite{Maldacena:2016hyu} addresses these obstacles:
\begin{itemize}
\item The strict AdS$_2$ theory is interpreted as an inert IR fixed point of a dual CFT$_1$. 

An interesting holographic theory is obtained only by perturbing away from the fixed point by irrelevant operators. These operators dominate the far UV, corresponding to the asymptotic AdS$_2$ boundary breaking down. However, their description of the approach to the IR is controlled. 

The spectrum we compute classifies the irrelevant operators in the IR fixed point theory that may serve as appropriate deformations. When these operators are added to the Lagrangian they deform the theory such that conformal symmetry is broken and new length scales are introduced. The most important scales appearing in this manner are associated with $h=2$ operators and were discussed in \cite{Larsen:2018iou}.
\item In constructions where AdS$_2$ arises from AdS$_3$ through a null reduction the dependence on the null direction indeed dominates in the strict infrared limit. However, the irrelevant operators controlling the {\it near} infrared regime are transverse to the direction of dimensional reduction and such excitations depend on position in the AdS$_2$ geometry. We identify our spectrum with such operators. 
\end{itemize}
In short, the spectrum given in table \ref{summary:h} does not describe the ground state of AdS$_2$ quantum gravity but rather the low lying excitations above the ground state. In terms of a CFT$_2$, the ground state has huge degeneracy and is referred to as left moving in our conventions. The nAdS$_2$ theory with the spectrum we compute characterizes the leading excitations which, for kinematic reasons, are entirely right moving and only weakly coupled to the left moving ground state. The discussion in section \ref{section:3D to 2D} elaborates on this interpretation and related conceptual challenges.

The simplifications we observe by explicit computations are, as mentioned, reminiscent of those that are due to supersymmetry. In section \ref{section:global symmetry} we develop this point of view and identify fermionic operators that generate the black hole spectra. It would be interesting to recover the same generators from {\it ab initio} considerations. Progress in this direction could yield clues to the microscopic description of these black holes. 

This paper is organized as follows. In section \ref{section:N=8} we describe the extremal non-BPS black hole backgrounds we consider as solutions to $\mathcal{N}=8$ (or $\mathcal{N}=4$) supergravity in $D=4$. They all have $AdS_2\times S^2$ near horizon geometry and in these contexts they respect $USp(8)$ (or $USp(4)\times SO(n_V-1)$) global symmetry. This symmetry structure partially diagonalizes the quadratic fluctuations around the backgrounds by organizing them into manageable blocks that are decoupled from one another. In section \ref{section:mass spectrum} we compute the mass spectrum of these blocks and obtain the conformal weights $h$ of the corresponding fields. In section \ref{section:heat kernels} we compute the logarithmic correction to the black hole entropy due the one loop contributions of all these states and find agreement with the results recently found using very different methods \cite{Castro:2018hsc}. In section \ref{section:3D to 2D}, we study the dimensional reduction from AdS$_3\times S^2$ to AdS$_2\times S^2$ and show how, depending on a choice of chirality, we reproduce either the nonBPS spectrum or the BPS spectrum on AdS$_2\times S^2$. This not only yields yet another consistency check on our computations but, as we discuss, it also enlightens the relation between the nAdS$_2$/nCFT$_1$ correspondence and black holes in string theory. We finish in section \ref{section:global symmetry} with a discussion of broken supersymmetry.

 \section{Black Holes and Their Fluctuations}\label{section:N=8}

In this section we introduce the nonBPS black holes in ${\cal N}=8$ and ${\cal N}=4$ supergravity. We exploit symmetries to establish the partial decoupling of quadratic fluctuations around these backgrounds into blocks.

\subsection{The AdS$_2\times S^2$ Backgrounds in ${\cal N}=8$ Supergravity}
$\mathcal{N}=8$ supergravity in $D=4$ spacetime dimensions consists of one graviton, $8$ gravitini $\Psi_{\hat{\mu}A}$, $28$ $U(1)$ vector fields $A_{\hat{\mu}}^{AB}$, $56$ Majorana spinors $\Lambda_{ABC}$, and $70$ scalars $W_{ABCD}$. The hatted greek indices $\hat{\mu},\hat{\nu}=0,1,2,3$ denote 4D Lorentz indices and capital latin letters $A=1,...,8$ refer to the global $SU(8)_R$ symmetry of $\mathcal{N}=8$ SUGRA. The $SU(8)_R$ indices are fully antisymmetrized so
the graviton, gravitini, vectors, gaugini, and scalars transform in representations \textbf{1}, \textbf{8}, \textbf{28}, \textbf{56} and \textbf{70}
of the $SU(8)_R$ group.

The black hole backgrounds we consider all have an AdS$_{2}\times S^{2}$ near horizon geometry,
 \begin{eqnarray}
{R}_{\mu\nu\lambda\rho}&=&-\frac{1}{\ell^2}({g}_{\mu\lambda}{g}_{\nu\rho}-{g}_{\mu\rho}{g}_{\nu\lambda})~,\\
{R}_{\alpha\beta\gamma\delta}&=&+\frac{1}{\ell^2}({g}_{\alpha\gamma}{g}_{\beta\delta}-{g}_{\alpha\delta}{g}_{\beta\gamma})~,
\end{eqnarray}
where unhatted indices $\mu,\nu=0,1$ and $\alpha,\beta=2,3$ refer to AdS$_{2}$ and $S^2$, respectively. $\ell$ is the radius of curvature of both 2D spaces.

The scalar fields are all constant on AdS$_{2}\times S^{2}$ and the fermions vanish. Thus the only matter supporting the geometry is the $28$ field strengths $G_{{\hat{\mu}} {\hat{\nu}}}^{AB} = 2\partial_{[{\hat{\mu}}} A_{{\hat{\nu}}]}^{AB}$. The $28$ electric charges (field components on AdS$_2$) and $28$ magnetic charges (field components on $S^2$) characterizing the field strengths can famously be organized into a fundamental representation {\bf 56} of $E_{7(7)}$ duality symmetry \cite{Cremmer:1979up}. However,
it is convenient to focus on the $SU(8)_R$ symmetry that is the maximal compact subgroup of $E_{7(7)}$ and express the charges by the complex antisymmetric central charge matrix $Z_{AB}$. After block diagonalization by an $SU(8)_R$ transformation we can present it as
\begin{eqnarray}
Z_{AB}=\text{diag}(\lambda_1\epsilon,\lambda_2\epsilon,\lambda_3\epsilon,\lambda_4\epsilon) ~, \quad
\epsilon=\begin{pmatrix}0&1\\-1&0\end{pmatrix}~.
\label{eqn:zabdef}
\end{eqnarray}

The canonical example of a charge configuration that corresponds to a BPS solution is $\lambda_1=\ell^{-1}$ and $\lambda_2=\lambda_3=\lambda_4=0$. These skew-eigenvalues preserve a $SU(2)_R\times SU(6)$ subgroup of $SU(8)_R$. The symmetry breaking pattern $SU(8)_R\to SU(2)_R\times SU(6)$ constitutes a more general characterization of the charges corresponding to BPS black holes with finite area.

A charge configuration that corresponds to the nonBPS black holes we focus on is \cite{Ferrara:2006em,Ceresole:2009jc}
\begin{eqnarray}
\lambda_1=\lambda_2=\lambda_3=\lambda_4=\frac{e^{i\frac{\pi}{4}}}{2\ell}~.
\label{eqn:nonbpslambdas}
\end{eqnarray}
It realizes the symmetry breaking pattern $SU(8)_R\to USp(8)$ that is characteristic of the nonBPS branch. To see this, note that
the central charge matrix \eqref{eqn:zabdef} with the skew-eigenvalues \eqref{eqn:nonbpslambdas} is proportional to the symplectic matrix
\begin{equation}
\Omega_8 = \text{diag}(\epsilon,\epsilon,\epsilon,\epsilon)~.
\label{eqn:omega8def}
\end{equation}
The antisymmetric tensor representation of $USp(8)$ is inherited from that of $SU(8)$ by imposing tracelessness upon contraction with $\Omega_8$ so the symmetry breaking $SU(8)_R\to USp(8)$ is manifest.

The phase appearing in \eqref{eqn:nonbpslambdas} ensures that the central charge matrix $Z_{AB}$ has determinant $+1$, as it must to be an element of $SU(8)_R$. Physically, the phase shows that the nonBPS branch has equal electric and magnetic charges, in contrast to the BPS solutions that can be chosen to have only electric charge. The factor $\frac{1}{2}$ on the right hand side of \eqref{eqn:nonbpslambdas} is such that the quadratic invariant $Z_{AB} Z^{AB}$ has the same magnitude for BPS and nonBPS black holes. This means the energy momentum tensor will be the same on the two branches which show that they share the same geometry.

In contrast, fermions enjoy Pauli couplings that depend linearly on the field strengths so supersymmetry acts differently on the two branches. Supersymmetry is preserved when the fermion transformations
\begin{eqnarray}
&&\delta \lambda^{ABC} =  - \frac{3}{\sqrt{2}} {\hat G}^{[AB} \epsilon^{C]}~,\\
&&\delta \psi^A_{\hat{\mu}} = \left(\delta^{AB} D_{\hat{\mu}} + \frac{1}{2} {\hat G}^{AB}\Gamma_{\hat{\mu}} \right) \epsilon_B~,
\label{eqn:deltapsiab}
\end{eqnarray}
vanish, where the field strengths ${\hat G}^{AB}\equiv\frac{1}{2}\Gamma^{\hat{\mu}\hat{\nu}}G^{AB}_{\hat{\mu}\hat{\nu}}$. We can assume without loss of generality that ${\hat G}^{AB}$ are block diagonal in the $(AB)$ indices, as for the central charge in \eqref{eqn:zabdef}. Thus the 4 sectors $(12),(34),(56),(78)$ do not couple to each other. On the BPS branch only ${\hat G}^{12}$ is nonvanishing. In this case there are no solutions for $\epsilon_B$ in the $(34), (56), (78)$ sectors but, in the $(12)$ sector, there is a solution with nontrivial $\epsilon_{1,2}$ and so the BPS solutions preserve ${\cal N}=2$ supersymmetry. On the nonBPS branch the $(12), (34), (56), (78)$ sectors give equivalent conditions but, because of the factor $\frac{1}{2}$ in \eqref{eqn:nonbpslambdas} that was discussed in the preceding paragraph, there is a mismatch between the magnitude of the field strength and the AdS$_2$ with scale $\ell$. Therefore, there are no solutions for $\epsilon_B$ on the nonBPS branch.

\subsection{Adaptation to ${\cal N}=4$ Supergravity}
We also want to discuss the spectrum of nonBPS black holes in ${\cal N}=4$ supergravity. It will ultimately follow automatically from the results in ${\cal N}=8$ supergravity, after a few modest reinterpretations.

In order to show this we first truncate ${\cal N}=8$ supergravity to ${\cal N}=4$ supergravity. This truncation
breaks the global symmetry $SU(8)_R \rightarrow SU(4)_R\times SU(4)_{\text{matter}}$. The branching rules of this symmetry breaking are
\begin{eqnarray}
{\bf 70} &\to & 2({\bf 1},{\bf 1})\oplus ({\bf 6},{\bf 6})\oplus ({\bf 4},{\bf {\bar 4}})\oplus ({\bf {\bar 4}}, {\bf 4}) ~, \cr
{\bf 56} &\to & ({\bf {\bar 4}},{\bf 1})\oplus ({\bf 6},{\bf 4})\oplus ({\bf 4}, {\bf 6}) \oplus ({\bf 1}, {\bf {\bar 4}}) ~, \cr
{\bf 28} &\to & ({\bf 1},{\bf 6})\oplus ({\bf 6},{\bf 1})\oplus ({\bf 4},{\bf 4}) ~, \cr
{\bf 8} &\to & ({\bf 1},{\bf 4})\oplus ({\bf 4},{\bf 1}) ~, \cr
{\bf 1} &\to & ({\bf 1},{\bf 1}) ~.
\label{eqn:bosetruncation}
\end{eqnarray}
It is a consistent truncation that preserves $\mathcal{N}=4$ supersymmetry to omit all fields in the ${\bf 4}$ (or ${\bf {\bar 4}}$) of
$SU(4)_{\text{matter}}$. The truncated theory obtained this way comprises an $\mathcal{N}=4$ supergravity multiplet (in the ${\bf 1}$ of $SU(4)_{\text{matter}}$) and $n_{V}=6$ matter multiplets (in the ${\bf 6}$ of $SU(4)_{\text{matter}}$).

The matter supporting AdS$_{2}\times S^{2}$ solutions in $\mathcal{N}=8$ supergravity is encoded in the spacetime central
charges \eqref{eqn:zabdef}. The nontrivial fields can be chosen without loss of generality as the four skew-diagonal ones and these are all retained in the truncation of $SU(8)_R$ to its $SU(4)_R\times SU(4)_{\text{matter}}$ subgroup. Therefore these background configurations are also solutions to the truncated theory with $\mathcal{N}=4$ supersymmetry. We focus on the nonBPS branch with skew-eigenvalues \eqref{eqn:nonbpslambdas} and the symmetry breaking pattern $SU(8)_R\rightarrow USp(8)$ in $\mathcal{N}=8$ SUGRA. This case descends to a nonBPS branch of $\mathcal{N}=4$ SUGRA
with the symmetry breaking pattern $SU(4)_R\times SU(4)_{\text{matter}}\rightarrow USp(4)\times USp(4)_{\text{matter}}$.

There is a simple generalization of this result to $\mathcal{N}=4$ SUGRA with a general number $n_V\geq1$ of matter multiplets \cite{Castro:2018hsc}.
Since $SU(4)=SO(6)$ and $USp(4)=SO(5)$ as Lie algebras, the symmetry breaking pattern of the nonBPS branch found in the preceding paragraph for $n_V=6$ matter multiplets is equivalent to $SO(n_V)_\text{matter}\rightarrow SO(n_V-1)_\text{matter}$. This is the pattern that characterizes the nonBPS solutions of theories with any $n_V\geq1$.

\subsection{Structure of Fluctuations}
As we have stressed, our background solution breaks the global $SU(8)_R$ of $\mathcal{N}=8$ SUGRA theory to a $USp(8)$ subgroup. This greatly simplifies the analysis of fluctuations around the background because it shows that different $USp(8)$ representations cannot couple at quadratic order. We can therefore organize the spectrum as representations of $USp(8)$.

The branchings of $SU(8)_R\to USp(8)$ for the matter representations in $\mathcal{N}=8$ SUGRA can be realized explicitly by removing contractions with the symplectic invariant \eqref{eqn:omega8def} from $SU(8)_R$ representations. This gives
\begin{eqnarray}
{\bf 70} &\to & {\bf 42}\oplus {\bf 27} \oplus {\bf 1}  ~, \cr
{\bf 56} &\to & {\bf 48} \oplus {\bf 8} ~, \cr
{\bf 28} &\to & {\bf 27} \oplus {\bf 1} ~, \cr
{\bf 8} &\to & {\bf 8} ~, \cr
{\bf 1} &\to & {\bf 1} ~.
\label{eqn:su8branching}
\end{eqnarray}
Collecting all singlets we find that on the nonBPS branch gravity can mix with one linear combination of the vector fields and similarly with one scalar. This is the field content of minimal Kaluza-Klein gravity in 4D, obtained by dimensional reduction of Einstein gravity in 5D. Truncation of $\mathcal{N}=8$ SUGRA to this sector is consistent and identifies the black holes on the nonBPS branch with the black holes in
Kaluza-Klein theory \cite{Rasheed:1995zv,Larsen:1999pp}. Moreover, the quadratic fluctuations of these fields is identical whether we consider the nonBPS branch of $\mathcal{N}=8$ SUGRA or minimal Kaluza-Klein theory. We therefore refer to the singlet sector as the ``Kaluza-Klein block".

The other $USp(8)$ representations similarly present ``blocks" that do not mix with each other. We summarize these decoupled sectors in table \ref{FluctTable}. The partial diagonalization of quadratic fluctuations into blocks was previously established away from extremality \cite{Castro:2018hsc}.

The spectrum of the KK black hole in $\mathcal{N}=4$ SUGRA can be computed directly, or by truncating the fluctuations analyzed for $\mathcal{N}=8$ SUGRA. The blocks of decoupled quadratic fluctuations are unchanged, it is only their degeneracy that is modified. Table \ref{FluctTable} lists the multiplicity of block in $\mathcal{N}=4$ SUGRA with $n_V\geq1$ matter multiplets and their representations under the global $USp(4)\times SO(n_V-1)_\text{matter}$ symmetry.

\renewcommand{\arraystretch}{2}
    \begin{table}[htbp]
	\centering
    \scalebox{0.85}[0.85]{
	\begin{tabular}{|c|c|c|c|c|c|c|}
		\hline
		\multirow{2}{*}{Multiplet}    & \multirow{2}{*}{Block content}   & \multirow{2}{*}{d.o.f.}  & \multicolumn{2}{c|}{$\mathcal{N}=8$} & \multicolumn{2}{c|}{$\mathcal{N}=4$ with $n_V$ matter multiplets}            \\ \cline{4-7}
		& & & $USp(8)$  & \#       & $USp(4)\times SO(n_V-1)_{\text{matter}}$ & \#\\ \hline\hline
		 KK block & 1 graviton, 1 vector, 1 scalar & 5 & {\bf 1} & 1 & ({\bf 1},{\bf 1})  & $1$\\\hline
		Gravitino block &2 gravitini and 2 gaugini & 8 & {\bf 8} & 4 & ({\bf 4},{\bf 1}) & $2$\\\hline
		Vector block  & 1 vector and 1 (pseudo)scalar & 3 &  {\bf 27} & 27 & ({\bf 5},{\bf 1}) $\oplus$ ({\bf 1},$n_V-1$) $\oplus$ ({\bf 1},{\bf 1}) & $n_V+5$\\\hline
		Gaugino block &2 gaugini & 4 & {\bf 48} & 24 & ({\bf 4},$n_V-1$) $\oplus$ ({\bf 4},{\bf 1})  & $2n_V$\\
		\hline
				Scalar block & 1 real scalar & 1 &  {\bf 42} & 42 & ({\bf 5},$n_V-1$) $\oplus$ ({\bf 1},{\bf 1}) & $5n_V-4$\\ \hline

	\end{tabular}}
    	\caption{Decoupled quadratic fluctuations around the KK black hole in $\mathcal{N}=8$ and $\mathcal{N}=4$ supergravity. The columns \# denote the multiplicity of the blocks.}\label{FluctTable}
\end{table}
\renewcommand{\arraystretch}{1}
    
\section{Mass Spectrum}
\label{section:mass spectrum}
In this section we compute the mass spectrum of fields on AdS$_2\times S^2$. Global symmetries partially decouple the fluctuations so we can consider one block at a time, as discussed in section \ref{section:N=8} and summarized in table \ref{FluctTable}.
For each block we start from the linearized equations of motion in 4D and expand the perturbations in spherical harmonics on $S^{2}$, before diagonalizing the resulting 2D equations of motion explicitly.  Bulk modes are analyzed in section \ref{section:bulk:scalar} through \ref{section:bulk:gravitino} and boundary modes are considered in section \ref{section:bndy}. From now on, we set the AdS$_2$ radius $\ell$ to $1$ for simplicity.

This section is long and relatively technical. Readers who are not interested in the detailed computations can jump directly to section \ref{section:spectrum summary} where the results are summarized.
    
\subsection{Partial Wave Expansion on $S^{2}$ and Dualization of AdS$_2$ Vectors}
The standard basis elements for the partial wave expansion of a scalar field on $S^{2}$ are the spherical harmonics $Y_{(lm)}$, i.e. the eigenfunctions of the 2D Laplacian $\nabla^2_S$ on $S^2$ with eigenvalues $-l(l+1)$. The analogous spherical harmonics for vector (or tensor) fields on $S^2$ are easily formed by taking one (or two) derivatives of $Y_{(lm)}$ along the $S^2$. 
Thus we can expand a 4D scalar $w$, a 4D vector $a_{\hat{\mu}}$, and 4D gravity $h_{\hat{\mu}\hat{\nu}}$ as
\begin{eqnarray}
\label{Expansion:b6} w&=&\sum_{lm}\varphi^{(lm)}Y_{(lm)}~,\\
\label{Expansion:b4}a_{\mu}&=&\sum_{lm}b^{(lm)}_{\mu}Y_{(lm)}~,\\
\label{Expansion:b5} a_{\alpha}&=&\sum_{lm}\left(b^{(lm)}_{1}\nabla_{\alpha}Y_{(lm)}+b^{(lm)}_{2}\epsilon_{\alpha\beta}\nabla^{\beta}Y_{(lm)}\right)~,\\
\label{Expansion:b1}
h_{\mu\nu}&=&\sum_{lm}H^{(lm)}_{\mu\nu}Y_{(lm)}~,\\    
\label{Expansion:b2}h_{\mu\alpha}&=&\sum_{lm}\left(B^{(lm)}_{1\mu}\nabla_{\alpha}Y_{(lm)}+B^{(lm)}_{2\mu}\epsilon_{\alpha\beta}\nabla^{\beta}Y_{(lm)}\right)~,\\
 \label{Expansion:b3}h_{\alpha\beta}&=&\sum_{lm}\left(\phi^{(lm)}_{1}\nabla_{\{\alpha}\nabla_{\beta\}}Y_{(lm)}+\phi^{(lm)}_{2}\epsilon_{\{\alpha}^{\quad\gamma}\nabla_{\beta\}}\nabla_{\gamma}Y_{(lm)}+\phi^{(lm)}_{3}g_{\alpha\beta}Y_{(lm)}\right)~.
\end{eqnarray}
Curly brackets indicate traceless symmetrization of indices such as $\nabla_{\{\alpha} \nabla_{\beta\}} = 
\frac{1}{2} ( \nabla_{\alpha}\nabla_{\beta} + \nabla_{\beta}\nabla_{\alpha}- g_{{\alpha}{\beta}}\nabla^2)$. The coefficient functions    
$H^{(lm)}_{\mu\nu}, B^{(lm)}_{1\mu},\ldots $ are fields on the AdS$_2$ base with AdS$_2$ tensor structure given by the indices $\mu, \nu,\ldots$ and degeneracy enumerated by the angular momentum quantum numbers $(lm)$.
    
Fermion fields can similarly be expanded on a basis of spinor spherical harmonics $\eta_{(\sigma lm)}$ satisfying $\gamma^{\alpha}D_{\alpha}\eta_{(\sigma lm)}=i(l+1)\eta_{(\sigma lm)}$ where $\gamma^{\alpha}$ denotes gamma matrices on $S^2$. We will use $\gamma^{\mu}$ for gamma matrices on AdS$_2$ and $\Gamma^{\hat{\mu}}$ for 4D gamma matrices. The partial wave expansion of a gaugino $\Lambda$ and a gravitino $\Psi_{{\hat\mu}}$ are
\begin{eqnarray}
\label{Expansion:f3} 
\Lambda &=&\lambda^{(\sigma lm)}_{+}\otimes \eta_{(\sigma lm)}+\lambda^{(\sigma lm)}_{-}\otimes\gamma_{S} \eta_{(\sigma lm)}~,\\
\label{Expansion:f1} 
\Psi_{\mu}&=&\psi^{(\sigma lm)}_{\mu+}\otimes \eta_{(\sigma lm)}+\psi^{(\sigma lm)}_{\mu-}\otimes\gamma_{S} \eta_{(\sigma lm)}~,\\
\label{Expansion:f2} 
\Psi_{\alpha}&=&\psi^{(\sigma lm)}_{+}\otimes D_{(\alpha)}\eta_{(\sigma lm)}+\psi^{(\sigma lm)}_{-}\otimes D_{(\alpha)}\gamma_{S}\eta_{(\sigma lm)}\nonumber\\
&&+\chi^{(\sigma lm)}_{+}\otimes\gamma_{\alpha}\eta_{(\sigma lm)}+\chi^{(\sigma lm)}_{-}\otimes\gamma_{\alpha}\gamma_{S}\eta_{(\sigma lm)}~,
\end{eqnarray}
where the summation symbol is suppressed for brevity. The chirality operator $\gamma_S$ is the $S^2$ analogue of $\Gamma_5$ in 4D and the symbol $D_{(\alpha)}=D_\alpha - \frac{1}{2}\gamma_\alpha \gamma^\beta D_\beta$. The indices $\pm$ on the fields on AdS$_2$ thus refer to chirality and the four terms in \eqref{Expansion:f3} correspond to projection on to the four helicities, $\pm\frac{3}{2}, \pm\frac{1}{2}$. There is a detailed discussion of spinors on $S^2$ in \cite{Abrikosov:2001nj}.

It will be sufficient to discuss bulk modes on-shell. Therefore, we can impose gauge conditions from the outset. The Lorentz-deDonder (LdD) gauge 
\begin{eqnarray}
\label{eqn:LdD}
\nabla^{\alpha}h_{\alpha\mu}=\nabla^{\alpha} h_{\{\alpha\beta\}} =0~,\qquad \nabla^{\alpha}a_{\alpha}=0~, \qquad \gamma^{\alpha}\Psi_{\alpha}=0~,
\end{eqnarray}
amounts to the conditions on AdS$_2$ fields
\begin{eqnarray}
&&\phi^{(lm)}_{1}=\phi^{(lm)}_{2}=0~,\quad B^{(lm)}_{1\mu}=0~,
\label{eqn:g1=0}
\\
&& b^{(lm)}_{1}=0~,
\label{eqn:b1=0}\\
&&\chi^{(\sigma lm)}_{+}=\chi^{(\sigma lm)}_{-}=0~.
    \end{eqnarray}
This simplifies the expansions (\ref{Expansion:b5}, \ref{Expansion:b2}, \ref{Expansion:b3}, \ref{Expansion:f2}). Importantly, the LdD gauge \eqref{eqn:LdD} is complete only for partial waves with $l\geq2$. For $l=0,1$ some of the LdD gauge conditions are vacuous so additional gauge fixing is needed. 
We will discuss this on a case by case basis. 

A vector field in AdS$_2$ can be dualized to two scalars as
\begin{eqnarray}
\label{Dualize:b}
&&b^{(lm)}_{\mu}=\epsilon_{\mu\nu}\nabla^{\nu}a^{(lm)}_{\perp}+\nabla_{\mu}a^{(lm)}_{\parallel}~.
\end{eqnarray}
This decomposition into transverse and longitudinal modes is unique when there are no normalizable harmonic scalars, as in Euclidean AdS$_2$. In Lorentzian signature there are nontrivial harmonic modes but they are not physical as they can be presented in longitudinal form where they manifestly decouple from physical processes. 
The determination of boundary modes in section \ref{section:bndy} will further refine these statements 
by considering nonnormalizable harmonic modes.

\subsection{Bulk Modes of the Scalar Block}
\label{section:bulk:scalar}

The scalar block consists of just one 4D scalar that is minimally coupled. Upon expansion in partial waves following \eqref{Expansion:b6}, the 4D Klein-Gordon equation becomes
\begin{eqnarray}
&&\left(\nabla^{2}_{A}-l(l+1)\right)\varphi^{(lm)}=0~,\quad l\geq0~.
\end{eqnarray}
The effective 2D mass is therefore $m^2=l(l+1)=j(j+1)$ after identification of the orbital angular momentum $l$ with the total angular momentum $j$, as usual for scalar fields. Therefore \eqref{eqn:hform} gives the conformal weight
\begin{eqnarray}
h=j+1~.
\end{eqnarray}
This result applies for all integral $j\geq 0$.

\subsection{Bulk Modes of the Vector Block}

The 4D vector block couples a scalar field $x$ and a gauge field through the Lagrangian \cite{Castro:2018hsc}
\begin{eqnarray}\label{EoM:block:v}
e^{-1}\mathcal{L}_{\text{vector}}=-\frac{1}{2}\nabla_{\hat{\mu}}x\nabla^{\hat{\mu}}x-\frac{1}{4}f_{\hat{\mu}\hat{\nu}}f^{\hat{\mu}\hat{\nu}}+xf_{\hat{\mu}\hat{\nu}}G^{\hat{\mu}\hat{\nu}}~,
\end{eqnarray}
where the background gauge field $G^{\hat{\mu}\hat{\nu}}$ has AdS$_2$ and $S^2$ components $G^{\mu\nu}=\frac{1}{\sqrt{2}}\epsilon^{\mu\nu}$ and $G^{\alpha\beta}=\frac{1}{\sqrt{2}}\epsilon^{\alpha\beta}$. The resulting 4D equations of motion for the scalar and the vector are
\begin{eqnarray}
&&\nabla^{2}x+f_{\hat{\mu}\hat{\nu}}G^{\hat{\mu}\hat{\nu}}=0~,\\
&&\nabla^{\hat{\mu}}\left(f_{\hat{\mu}\hat{\nu}}-2xG^{\hat{\mu}\hat{\nu}}\right)=0~.
\end{eqnarray}
Applying partial wave expansions of the form (\ref{Expansion:b4}, \ref{Expansion:b5}), the gauge condition \eqref{eqn:b1=0}, and dualization \eqref{Dualize:b} we find
\begin{eqnarray}
\label{EoM_V:s}&&\left[\left(\nabla^{2}_{A}-l(l+1)\right)x+\sqrt{2}l(l+1)b_{2}+\sqrt{2}\nabla^{2}_{A}a_{\perp}\right]Y=0~,\\
\label{EoM_V:vmu}&&\epsilon_{\nu\mu}\nabla^{\mu}\left[\left(\nabla^{2}_{A}-l(l+1)\right)a_{\perp}+\sqrt{2}x\right]Y+\nabla_{\nu}\left[\left(\nabla^{2}_{A}-l(l+1)\right)a_{\parallel}\right]Y=0~,\\
\label{EoM_V:va}&&\left[-\nabla^2_{A}a_{\parallel}\right]\nabla_{\alpha}Y+\left[\left(\nabla^{2}_{A}-l(l+1)\right)b_{2}+\sqrt{2}x\right]\epsilon_{\alpha\beta}\nabla^{\beta}Y=0~.
\end{eqnarray}
The partial wave numbers $(lm)$ on the 2D fields $x, b_2, a_{\perp}, a_{\parallel}$ and on the spherical harmonics $Y$ are suppressed for brevity. Since $Y_{(00)}$ is a constant on $S^2$ \eqref{EoM_V:va} has no content for $l=0$. For the same reason, the expansion \eqref{Expansion:b5} in vector harmonics on $S^2$ leaves the component $b^{(00)}_2$ undefined. Importantly, the combination $l(l+1)b^{(00)}_2$ unambiguously 
vanishes for $l=0$, so \eqref{EoM_V:s} is meaningful for all $l\geq 0$. 

The 4D equations of motion (\ref{EoM_V:s}, \ref{EoM_V:vmu}, \ref{EoM_V:va}) are equivalent to the vanishing of each expression in square bracket by itself, due to orthogonality of the spherical harmonics.  For \eqref{EoM_V:vmu} we also appeal to uniqueness of dualization in order to remove the gradients on AdS$_2$. In the following we diagonalize these 2D equations of motion. We first discuss modes with $l\geq1$ and then address the special case $l=0$. 

\bigskip
\noindent
{\bf \emph{Vector block: $l\geq1$ modes} }

For $l\geq1$ we can apply \eqref{EoM_V:va}. In particular, the first equation shows that $a_{\parallel}=0$, due to the absence of propagating harmonic modes. Then (\ref{EoM_V:s}, \ref{EoM_V:vmu}, \ref{EoM_V:va}) give
\begin{eqnarray}
&&\left(\nabla^{2}_{A}-l(l+1)\right)x+\sqrt{2}l(l+1)b_{2}+\sqrt{2}\nabla^{2}_{A}a_{\perp}=0~,\\
&&\left(\nabla^{2}_{A}-l(l+1)\right)a_{\perp}+\sqrt{2}x=0~,\\
&&\left(\nabla^{2}_{A}-l(l+1)\right)b_{2}+\sqrt{2}x=0~,
\end{eqnarray}
which can be reordered into the diagonal form
\begin{eqnarray}
&&\left(\nabla^{2}_{A}-(l-1)l\right)\left(\sqrt{2}x+(l+1)(a_{\perp}+b_{2})\right)=0~,\\
&&\left(\nabla^{2}_{A}-l(l+1)\right)\left(a_{\perp}-b_{2}\right)=0~,\\
&&\left(\nabla^{2}_{A}-(l+1)(l+2)\right)\left(\sqrt{2}x-l(a_{\perp}+b_{2})\right)=0~,
\end{eqnarray}
The eigenvalues of the AdS$_2$ Laplacian $\nabla^{2}_{A}$ thus give the scalar masses
\begin{eqnarray}
&&m^{2}=(l-1)l~,\ l(l+1)~,\ (l+1)(l+2)~,
\end{eqnarray}
and so the conformal weights \eqref{eqn:hform} become
\begin{eqnarray}
\label{eqn:vecjgeq1}
h=j ~,\ j+1~,\ j+2~,
\end{eqnarray}
for all integral $j\geq1$. We identified the angular quantum number $j=l$ by noting that each value of the conformal weight has degeneracy $(2l+1)$, the dimension of the irreducible representation of $SU(2)$ with $j=l$.

\bigskip
\noindent
{\bf \emph{Vector block: $l=0$ modes} }
	
In the $l=0$ sector  The 4D gauge field $b_{\hat\mu}$ has no components on the $S^2$ so the only non-vanishing field components are $a_{\perp}$, $a_{\parallel}$, and $x$.
Since $Y_{(00)}=1$ the LdD gauge condition \eqref{eqn:LdD} is empty for $l=0$. On the other hand, the standard 4D gauge transformation 
$b_{\hat\mu} \to b_{\hat\mu} + \partial_{\hat\mu} \Lambda$ reduces to a 2D symmetry acting on the AdS$_2$ components $b_\mu$ because for $l=0$ it does not act on the (non-existent) components $b_\alpha$ of the vector field on $S^2$. We can exploit this gauge symmetry to set the longitudinal component $a_{\parallel}=0$.
The equations of motion (\ref{EoM_V:s}, \ref{EoM_V:vmu}) then give
\begin{eqnarray}\label{EoM_V:l=0}
\begin{cases}
\nabla^{2}_{A}x+\sqrt{2}\nabla^{2}_{A}a_{\perp}=0\\
\nabla^{2}_{A}a_{\perp}+\sqrt{2}x=0
\end{cases}\Rightarrow\quad
\begin{cases}
\left(\nabla^{2}_{A}-2\right)x=0\\
\nabla^{2}_{A}\left(\sqrt{2}a_{\perp}+x\right)=0
\end{cases}~.
\end{eqnarray}

The lower equation becomes a constraint $\sqrt{2}a_{\perp}+x=0$ up to a harmonic solution for $a_\perp$ which is equivalent to $a_{\parallel}$ that vanishes due to the gauge condition. The $l=0$ sector therefore reduces to one degree of freedom which we can identify as the scalar field $x$. This is the expected result because the vector block consists of a scalar and a vector but 2D vector fields have no degrees of freedom.

The upper equation in \eqref{EoM_V:l=0} identifies the eigenvalue of the scalar as $m^2=2$ which corresponds to conformal weight $h=2$. 
We can present this in terms of the result \eqref{eqn:vecjgeq1} for $j\geq 1$: the tower with $h=j+2$ is 
completed so it includes an entry for $j=0$ while the other two towers have no $j=0$ mode.

\subsection{Bulk Modes of the KK Block}\label{section:bulk:KK}
Expansion of the Kaluza-Klein Lagrangian to quadratic order around the AdS$_2\times S^2$ background supported by nonBPS fluxes yields a Lagrangian for the quadratic fluctuations (given explicitly in \cite{Castro:2018hsc}). This in turn gives the equations of motion for the KK block, summarized in the following.

\bigskip
\noindent
{\bf \emph{KK block: Einstein equation} }

The 4D Einstein equation is given by
\begin{eqnarray}
\label{EoM:Einstein}
&&\nabla^{2} h_{\hat{\mu}\hat{\nu}}+\nabla_{\hat{\mu}}\nabla_{\hat{\nu}}h-2\nabla_{(\hat{\mu}}\nabla^{\hat{\alpha}}h_{\hat{\nu})\hat{\alpha}}
-2{R}^{\hat{\alpha}}_{\ (\hat{\mu}}h_{\hat{\nu})\hat{\alpha}}-2{R}_{\hat{\alpha}\hat{\mu}\hat{\nu}\hat{\beta}}h^{\hat{\alpha}\hat{\beta}}+h_{\hat{\mu}\hat{\nu}}{R}\cr
&&+{g}_{\hat{\mu}\hat{\nu}}(-\nabla^{2}h+\nabla_{\hat{\alpha}}\nabla_{\hat{\beta}}h^{\hat{\alpha}\hat{\beta}}-h^{\hat{\alpha}\hat{\beta}}{R}_{\hat{\alpha}\hat{\beta}})=-8{G}_{(\hat{\mu}}^{\quad\hat{\alpha}}f_{\hat{\nu})\hat{\alpha}}+4{G}_{\hat{\mu}\hat{\alpha}}{G}_{\hat{\nu}\hat{\beta}}h^{\hat{\alpha}\hat{\beta}}
\cr
&&+2{g}_{\hat{\mu}\hat{\nu}}({G}^{\hat{\alpha}\hat{\beta}}f_{\hat{\alpha}\hat{\beta}}-{G}_{\hat{\alpha}}^{\ \hat{\nu}}{G}_{\hat{\beta}\hat{\nu}}h^{\hat{\alpha}\hat{\beta}})+h_{\hat{\mu}\hat{\nu}}{G}^{\hat{\alpha}\hat{\beta}}{G}_{\hat{\alpha}\hat{\beta}}+8\sqrt{3}\varphi{G}_{\hat{\mu}}^{\ \hat{\alpha}}{G}_{\hat{\nu}\hat{\alpha}}~.
\end{eqnarray}
The background is described by the 4D metric $g_{{\hat\mu}{\hat\nu}}$ with Riemann curvature $R^{{\hat\alpha}}_{\ {\hat\beta}{\hat\gamma}{\hat\delta}}$ as well as the gauge fields ${G}_{{\mu}{\nu}}=\frac{1}{\sqrt{2}}\epsilon_{{\mu}{\nu}}$ along AdS$_2$ and ${G}_{\alpha\beta}=\frac{1}{\sqrt{2}}\epsilon_{\alpha\beta}$ through $S^2$. The fluctuations are the metric $h_{{\hat\mu}{\hat\nu}}$, the field strength $f_{{\hat\mu}{\hat\nu}}$, and the scalar field $\varphi$. 

The partial wave expansions of the 4D fields take the form (\ref{Expansion:b6}-\ref{Expansion:b3}). 
Considering first the equations where $\hat{\mu}\hat{\nu}=\mu\nu$ so both indices are within AdS$_2$ we find
\begin{eqnarray}
\label{EoM:gmunu:trace}
&&\left[\left(l(l+1)+2\right) H-2\left(\nabla^{2}_{A}-l(l+1)\right)\phi_{3}+4\sqrt{2}\nabla^{2}_{A}a_{\perp}-4\sqrt{2}l(l+1)b_{2}+8\sqrt{3}\varphi\right]Y=0~,\\
\label{EoM:gmunu:traceless}
&&\left[-l(l+1)H_{\{\mu\nu\}}+2\nabla_{\{\mu}\nabla_{\nu\}}\phi_{3}\right]Y=0~,
\end{eqnarray}
for the scalar and symmetric traceless components of the AdS$_2$ indices $\mu\nu$. We suppress the partial wave indices $(lm)$ on the 2D fields to avoid clutter. The analogous equations for $\hat{\mu}\hat{\nu}=\alpha\beta$ so both indices of the Einstein equation \eqref{EoM:Einstein} are on the $S^2$ give 
\begin{eqnarray}
\label{EoM:gab}
&&\left[\nabla_{\rho}\nabla_{\sigma}H^{\rho\sigma}-\left(\nabla^{2}_{A}-\frac{1}{2}l(l+1)\right)H-\left(\nabla^{2}_{A}+2\right)\phi_{3}-2\sqrt{2}\nabla^{2}_{A}a_{\perp}+2\sqrt{2}l(l+1)b_{2}\right.\cr
&&\left.-4\sqrt{3}\varphi\right]g_{\alpha\beta}Y+\left[ H \right] ~\nabla_{\{\alpha}\nabla_{\beta\}}Y-\left[ 2\nabla^{2}_{A}B_{2\parallel}\right] ~\epsilon_{\{\alpha}^{\quad\gamma}\nabla_{\beta\}}\nabla_{\gamma}Y=0~.
\end{eqnarray}
Finally, the partial wave expansion of the Einstein equation with mixed indices $\hat{\mu}\hat{\nu}=\mu\alpha$ becomes
\begin{eqnarray}
\label{EoM:gmua}
&&\left(\epsilon_{\mu\nu}\nabla^{\nu}\left[(\nabla_{A}^{2}-l(l+1))B_{2\perp}-2\sqrt{2}b_{2}-2\sqrt{2}a_{\perp}\right]-\nabla_{\mu}\left[l(l+1)B_{2\parallel}+2\sqrt{2}a_{\parallel}\right]\right)\epsilon_{\alpha\beta}\nabla^{\beta}Y\nonumber\\
&&+\left[\nabla_{\mu}H-\nabla^{\nu}H_{\mu\nu}+\nabla_{\mu}\phi_{3}+2\sqrt{2}\epsilon_{\mu\nu}b^{\nu}-2\sqrt{2}\nabla_{\mu}b_{2}+2\epsilon_{\mu\nu}B^{\nu}_{2}\right]\nabla_{\alpha}Y=0~.
\end{eqnarray}

%

\bigskip
\noindent
{\bf \emph{KK block: vector equation} }
    
The equation of motion for the vector field in KK theory is
\begin{eqnarray}\label{EoM:Maxwell}
&&\nabla^{\hat{\mu}}\left(f_{\hat{\mu}\hat{\nu}}-h_{\hat{\mu}\hat{\rho}}{G}^{\hat{\rho}}_{\ \hat{\nu}}+h_{\hat{\nu}\hat{\rho}}{G}^{\hat{\rho}}_{\ \hat{\mu}}+\frac{1}{2}h{G}_{\hat{\mu}\hat{\nu}}-2\sqrt{3}\varphi {G}_{\hat{\mu}\hat{\nu}}\right)=0~,
\end{eqnarray}
after linearizing around our background. For $\hat{\nu}=\nu$ the 4D index is along AdS$_2$ and the partial wave expansions (\ref{Expansion:b6}-\ref{Expansion:b3}) give
\begin{eqnarray}\label{EoM:vmu}
&&\left(\epsilon_{{\nu}{\mu}}\nabla^{{\mu}}\left[\left(\nabla_{A}^{2}-l(l+1)\right)a_{\perp}-\frac{1}{\sqrt{2}}l(l+1)B_{2\perp}+\frac{1}{2\sqrt{2}}H-\frac{1}{\sqrt{2}}\phi_{3}+\sqrt{6}\varphi\right]\right.\cr
&&\quad\left.-\nabla_{\nu}\left[\nabla^{2}_{A}a_{\parallel}+\frac{1}{\sqrt{2}}l(l+1)B_{2\parallel}\right]\right)Y=0~.
\end{eqnarray}
We used the identity $\nabla_{{\mu}}H^{\{\mu\rho\}}\epsilon_{{\rho}{\nu}}=\nabla_{{\mu}}H_{\{\nu\rho\}}\epsilon^{{\rho}{\mu}}$. The partial wave expansion of the 4D field equation \eqref{EoM:Maxwell} for $\hat{\nu}=\alpha$ similarly gives
\begin{eqnarray}\label{EoM:va}
&&\left[\left(\nabla^{2}_{A}-l(l+1)\right)b_{2}+\frac{1}{\sqrt{2}}\phi_{3}-\frac{1}{\sqrt{2}}\nabla^{2}_{A}B_{2\perp}-\frac{1}{2\sqrt{2}}H+\sqrt{6}\varphi\right]\epsilon_{\alpha\beta}\nabla^{\beta}Y\cr
&&-\nabla^{2}_{A}\left[a_{\parallel}+\frac{1}{\sqrt{2}}B_{2\parallel}\right]\nabla_{\alpha}Y=0~.
 \end{eqnarray}
 %
 
\bigskip
\noindent
{\bf \emph{KK block: scalar equation} }
    
The last equation of motion for KK theory is the one for the KK scalar:
\begin{eqnarray}\label{EoM:Scalar}
&&8\nabla^{2}\varphi+8\sqrt{3}{G}^{\hat{\mu}\hat{\nu}}f_{\hat{\mu}\hat{\nu}}-4\sqrt{3}{R}^{\hat{\mu}\hat{\nu}}h_{\hat{\mu}\hat{\nu}}=0~.
\end{eqnarray}
The partial wave expansion gives
\begin{eqnarray}\label{EoM:s}
&&\left[\left(\nabla_{A}^{2}-l(l+1)\right)\varphi+\sqrt{6}\nabla^{2}_{A}a_{\perp}+\sqrt{6}l(l+1)b_{2}+\frac{\sqrt{3}}{2}H-\sqrt{3}\phi_{3}\right]Y=0~.
\end{eqnarray}

At this point we must solve all these equations. Orthogonality of spherical harmonics show that all terms in square brackets vanish. However, we must take into account that gradients $\nabla_\alpha Y$ of the spherical harmonics vanish for $l=0$ and traceless combinations of the double gradients $\nabla_\alpha \nabla_\beta Y$ vanish also for $l=1$. Therefore 
we first discuss the equations for $l\geq2$ and then address $l=1,0$. 

\bigskip
\noindent
{\bf \emph{KK block: $l\geq2$ modes} }
    
From (\ref{EoM:gab}) and (\ref{EoM:va}) we find
\begin{eqnarray}
&&\nabla^{2}_{A}B_{2\parallel}=\nabla^{2}_{A}a_{\parallel} =0~,\\
&&H=0~.
\end{eqnarray}
The uniqueness of AdS$_2$ dualization (up to modes that decouple) means we can take all these fields to vanish $B_{2\parallel}=a_{\parallel}=H=0$~. Additionally (\ref{EoM:gmunu:traceless}) 
shows that the graviton perturbations $H_{\{\mu\nu\}}$ can be expressed in terms of $\phi_3$ so they do not represent independent degrees of freedom. 

Taking these simplification into account, we gather the equations of motions 
(\ref{EoM:gmunu:trace}, \ref{EoM:gmua}, \ref{EoM:vmu}, \ref{EoM:va}, \ref{EoM:s}) and find
\begin{eqnarray}
&&\left(\nabla^{2}_{A}-l(l+1)\right)\phi_{3}=2\sqrt{2}\nabla^{2}_{A}a_{\perp}-2\sqrt{2}l(l+1)b_{2}+4\sqrt{3}\varphi~,\\
&&\left(\nabla_{A}^{2}-l(l+1)\right)B_{2\perp}=2\sqrt{2}b_{2}+2\sqrt{2}a_{\perp}~,\\
&&\left(\nabla_{A}^{2}-l(l+1)\right)a_{\perp}=\frac{1}{\sqrt{2}}l(l+1)B_{2\perp}+\frac{1}{\sqrt{2}}\phi_{3}-\sqrt{6}\varphi~,\\
&&\left(\nabla^{2}_{A}-l(l+1)\right)b_{2}=-\frac{1}{\sqrt{2}}\phi_{3}+\frac{1}{\sqrt{2}}\nabla^{2}_{A}B_{2\perp}-\sqrt{6}\varphi~,\\
&&\left(\nabla_{A}^{2}-l(l+1)\right)\varphi=-\sqrt{6}\nabla^{2}_{A}a_{\perp}-\sqrt{6}l(l+1)b_{2}+\sqrt{3}\phi_{3}~.
\end{eqnarray}
We can reorganize these equations as
\begin{eqnarray}
\label{eqn:gravl=2}
\label{eigen:(l+2)(l+3)}&&\left(\nabla^{2}_{A}-(l+2)(l+3)\right)\left[2\sqrt{3}\varphi-l(l+1)B_{2\perp}-2\sqrt{2}la_{\perp}-2\sqrt{2}lb_{2}\right]=0~,\\
\label{eigen:(l+1)(l+2)}&&\left(\nabla_{A}^{2}-(l+1)(l+2)\right)\left[-\phi_{3}-lB_{2\perp}-\sqrt{2}la_{\perp}+\sqrt{2}lb_{2}\right]=0~,\\
\label{eigen:l(l+1)}&&\left(\nabla_{A}^{2}-l(l+1)\right)\left[2\varphi+\sqrt{3}(l^{2}+l-1)B_{2\perp}+\sqrt{6}a_{\perp}+\sqrt{6}b_{2}\right]=0~,\\
\label{eigen:(l-1)1}&&\left(\nabla^{2}_{A}-(l-1)l\right)\left[\phi_{3}-(l+1)B_{2\perp}-\sqrt{2}(l+1)a_{\perp}+\sqrt{2}(l+1)b_{2}\right]=0~,\\
\label{eigen:(l-2)(l-1)}&&\left(\nabla_{A}^{2}-(l-2)(l-1)\right)\left[2\sqrt{3}\varphi-l(l+1)B_{2\perp}+2\sqrt{2}(l+1)\left(a_{\perp}+b_{2}\right)\right]=0~.
\end{eqnarray}
The scalar masses read off from the eigenvalues of $\nabla^{2}_{A}$ are
\begin{eqnarray}
m^{2}&=&(l-2)(l-1)~,\ (l-1)l~,\ l(l+1)~,\ (l+1)(l+2)~,\ (l+2)(l+3)~.
\end{eqnarray}
Each of the AdS$_2$ scalars have degeneracy $(2l+1)$ so we identify $j=l$, where $j$ is the angular quantum number labeling the irreducible representation of $SU(2)$. The conformal weights \eqref{eqn:hform} of the 1D conformal fields dual to the five partial wave towers of the KK block become
\begin{eqnarray}
\label{eqn:gravh}
h=~j-1~,\ j~,\ j+1~,\ j+2~,\ j+3~.
\end{eqnarray}
This result is valid for $j\geq2$.

\bigskip
\noindent
{\bf \emph{KK block: $l=1$ modes} }

The $l=1$ sector is special because $\epsilon_{\alpha\beta}\nabla^{\beta}Y_{(1m)}$/$\nabla_{\alpha}Y_{(1m)}$ are Killing Vectors (KVs)/Conformal Killing Vectors (CKVs) on $S^2$. Therefore $\epsilon_{\{\alpha}^{\quad\gamma}\nabla_{\beta\}}\nabla_{\gamma}Y_{(1m)}=\nabla_{\{\alpha}\nabla_{\beta\}}Y_{(1m)}=0$ and so the partial wave expansion \eqref{Expansion:b3} does not include the coefficient functions $\phi^{(1m)}_{1}$ and $\phi^{(1m)}_{2}$. Moreover, the gauge conditions $\nabla^{\alpha} h_{\{\alpha\beta\}}=0$ are automatic, they fail to constrain diffeomorphisms $\xi_\alpha$ on the $S^2$. 

We gauge fix the diffeomorphisms along the KVs by setting $B^{(1m)}_{2\parallel}=0$ and those along the CKVs by taking $\phi^{(1m)}_{3}=0$. 
With these conditions \eqref{EoM:gmunu:traceless} becomes a constraint that sets $H^{(1m)}_{\{\mu\nu\}}=0$ and the vanishing of the second square bracket in \eqref{EoM:gmua} demands that also $a^{(1m)}_{\parallel}=0$. 

After gauge fixing the $15$ partial wave components in the generic KK-block have been reduced to only $5$. We gather the remaining terms in (\ref{EoM:gmunu:trace}, \ref{EoM:gmua}, \ref{EoM:vmu}, \ref{EoM:va}, \ref{EoM:s}) for $l=1$ and get the equations of motion for these $5$ fields in AdS$_2$:
\begin{eqnarray}
&&H^{(1m)}=-\sqrt{2}\nabla^{2}_{A}a^{(1m)}_{\perp}+2\sqrt{2}b^{(1m)}_{2}-2\sqrt{3}\varphi^{(1m)}~,\\
&&\left(\nabla_{A}^{2}-2\right)B^{(1m)}_{2\perp}=2\sqrt{2}b^{(1m)}_{2}+2\sqrt{2}a^{(1m)}_{\perp}~,\\
&&\left(\nabla_{A}^{2}-2\right)a^{(1m)}_{\perp}=-\frac{1}{2\sqrt{2}}H^{(1m)}+\sqrt{2}B^{(1m)}_{2\perp}-\sqrt{6}\varphi^{(1m)}~,\\
&&\left(\nabla^{2}_{A}-2\right)b^{(1m)}_{2}=\frac{1}{\sqrt{2}}\nabla^{2}_{A}B^{(1m)}_{2\perp}+\frac{1}{2\sqrt{2}}H^{(1m)}-\sqrt{6}\varphi^{(1m)}~,\\
&&\left(\nabla_{A}^{2}-2\right)\varphi^{(1m)}=-\sqrt{6}\nabla^{2}_{A}a^{(1m)}_{\perp}-2\sqrt{6}b^{(1m)}_{2}-\frac{\sqrt{3}}{2}H^{(1m)}~.
\end{eqnarray}
Simplifying the first of these equations using the others we find
\begin{equation}
H^{(1m)}= -4\sqrt{2}a^{(1m)}_{\perp}+4\sqrt{2}b^{(1m)}_{2}-4B^{(1m)}_{2\perp} ~.
\end{equation}
Therefore $H^{(1m)}$ is not an independent field. We diagonalize the remaining equations as 
\begin{eqnarray}
\label{eqn:1gravl=1}
&&(\nabla^{2}_{A}-12)\left(-B^{(1)}_{2\perp}-\sqrt{2}a^{(1)}_{\perp}-\sqrt{2}b^{(1)}_{2}+\sqrt{3}\varphi^{(1)}\right)=0~,\\
\label{eqn:1gravl=2}
&&(\nabla^{2}_{A}-6)\left(-B^{(1)}_{2\perp}-\sqrt{2}a^{(1)}_{\perp}+\sqrt{2}b^{(1)}_{2}\right)=0~,\\
\label{eqn:1gravl=3}
&&(\nabla_{A}^{2}-2)\left(\sqrt{3}B^{(1)}_{2\perp}+\sqrt{6}a^{(1)}_{\perp}+\sqrt{6}b^{(1)}_{2}+2\varphi^{(1)}\right)=0~,\\
\label{l1:B:zero} &&\nabla^{2}_{A}\left(-B^{(1)}_{2\perp}+2\sqrt{2}a^{(1)}_{\perp}+2\sqrt{2}b^{(1)}_{2}+\sqrt{3}\varphi^{(1)}\right)=0~.
\end{eqnarray}
The final equation amounts to the constraint  
\begin{eqnarray}
&&-B^{(1)}_{2\perp}+2\sqrt{2}a^{(1)}_{\perp}+2\sqrt{2}b^{(1)}_{2}+\sqrt{3}\varphi^{(1)}=0~.
\end{eqnarray}
up to a harmonic function that can be fixed by residual symmetry. 

The three eigenvectors that remain represent propagating modes. This is the expected net number of physical fields from a gauge field and a scalar, the field content in the $l=1$ sector of the KK block. The source of all the complications addressed here is the mixing of these degrees of freedom with gravity and with each other. 
   
Therefore, for $j=l=1$, the eigenvalues of $\nabla^{2}_{A}$ are $m^2=12, 6, 2$, corresponding to the conformal weights $h=4,3,2$ respectively. 
Among the five towers in \eqref{eqn:gravh}, we thus find that those with $h=j+1,j+2,j+3$ are extended to $j=1$ while the towers with $h=j-1, j$ do not include modes $j=1$. Indeed, the three eigenvectors (\ref{eqn:1gravl=1}, \ref{eqn:1gravl=2}, \ref{eqn:1gravl=3}) with eigenvalues $12,6,2$ found for $l=1$ extend those 
identified in (\ref{eigen:(l+2)(l+3)}, \ref{eigen:(l+1)(l+2)}, \ref{eigen:l(l+1)}) for $l\geq 2$.

\bigskip
\noindent
{\bf \emph{KK block: $l=0$ modes} }
    
The spherical harmonic $Y_{(00)}=1$ is constant, so for $l=0$ the only non-vanishing terms defined by the partial wave expansions (\ref{Expansion:b6}-\ref{Expansion:b3}) are the 2D metric $H^{(00)}_{\mu\nu}$, the 2D gauge field $b^{(00)}_{\mu}$, the KK scalar $\varphi^{(00)}$ and the $S^2$ volume $\phi^{(00)}_{3}$. This is a total of $7$ non-vanishing 2D field components for $l=0$. In the $l=0$ sector the LdD gauge conditions \eqref{eqn:LdD} place no restrictions on the fields. The 2D diffeomorphism symmetry generated by an AdS$_2$ vector $\xi_\mu$ is therefore unfixed, as is the 2D gauge symmetry. We fix these three symmetries by imposing 
\begin{eqnarray}
\label{eqn:l=0cond}
&&\nabla^{\mu}\nabla^{\nu}H^{(00)}_{\{\mu\nu\}} = 0~,\qquad a^{(00)}_{\parallel}=0~.
\end{eqnarray}   
Notice that there are still residual diffeomorphisms that satisfy $\nabla^{\mu}\nabla^{\nu}\nabla_{\{\mu}\xi^{(00)}_{\nu\}} = 0$, which we will take advantage of later.

 The 2D equations of motion (\ref{EoM:gmunu:trace}, \ref{EoM:gmunu:traceless}, \ref{EoM:gab}, \ref{EoM:vmu}, \ref{EoM:s}) of the remaining $4$ field components $a^{(00)}_{\perp}$, $\phi^{(00)}_{3}$, $H^{(00)}$, and $\varphi^{(00)}$ can be organized as 
\begin{eqnarray}
\label{eqn:aperp}
&&\nabla_{A}^{2}\left(\sqrt{6}a^{(00)}_{\perp}+\varphi^{(00)}+\frac{\sqrt{3}}{4}H^{(00)} + 2\sqrt{3} \phi_3^{(00)} \right)=0~,\\
\label{eqn:phi3}
&&\nabla_{\{\mu}\nabla_{\nu\}}\phi^{(00)}_{3} = 0~,\\
&&(\nabla^{2}_{A}-2)\phi_3^{(00)}=0~,\\
\label{eqn:H0}
&&(\nabla^{2}_{A}-2)H^{(00)}= - 12 \phi_3^{(00)}~,\\
&&\left(\nabla_{A}^{2}-6\right)\varphi^{(00)}=0~.
\end{eqnarray}

Now \eqref{eqn:aperp} amounts to a constraint that expresses $a^{(00)}_{\perp}$ in terms of other fields, up to a harmonic mode that is inconsequential for the physical spectrum. Similarly, \eqref{eqn:phi3} define Conformal Killing Vectors (CKVs) $\nabla_\mu \phi^{(00)}_{3}$ but, since there are no normalizable CKVs on (Euclidean) AdS$_2$, we must have $\phi^{(00)}_{3}=0$. Then \eqref{eqn:H0} becomes 
\begin{eqnarray}
\label{eqn:H0:mass2}
(\nabla^{2}_{A}-2)H^{(00)}=0~.
\end{eqnarray}
However, the gauge conditions \eqref{eqn:l=0cond} permit residual diffeomorphisms $\xi_\mu$ satisfying
\begin{eqnarray}\label{eqn:H0:Dif}
&&\nabla^{\mu}\nabla^{\nu}\nabla_{\{\mu}\xi^{(00)}_{\nu\}} = 0 \quad \Leftrightarrow \quad (\nabla^{2}_{A}-2)\delta H^{(00)}=0~.
\end{eqnarray}
Such $\xi_\mu$ are CKVs that are necessarily nonnormalizable, but they correspond to normalizable $\delta H^{(00)}_{\mu\nu}$. Comparison of \eqref{eqn:H0:mass2} and \eqref{eqn:H0:Dif} shows that $H^{(00)}$ is pure gauge; it can be set to be zero by residual diffeomorphisms $\xi^{(00)}_{\mu}$.

In summary, in the $l=0$ sector of the KK-block there is only one physical degree of freedom which can be identified as $\varphi^{(00)}$. This mode generalizes the partial wave tower \eqref{eqn:gravl=2} to $l=0$. It is an eigenfunction of $\nabla_{A}^{2}$ with eigenvalue $m^2=6$, corresponding to $h=3$. Thus it extends the final entry $h=j+3$ in \eqref{eqn:gravh} to the value $j=0$. 

\subsection{Bulk Modes of the Gaugino Block}
\label{section:bulk:gaugino}
The gaugino block has the 4D Lagrangian \cite{Castro:2018hsc}
\begin{eqnarray}
&&e^{-1}\delta^{2}\mathcal{L}_{\text{gaugino}}=-\bar{\Lambda}_{A}\Gamma^{\hat{\mu}}D_{\hat{\mu}}\Lambda_{A}-\frac{1}{2}\epsilon_{AB}\bar{\Lambda}_{A}\hat{G}\Lambda_{B}~,
\end{eqnarray}
where $\hat{G}\equiv\frac{1}{2}\Gamma^{\hat{\mu}\hat{\nu}}G_{\hat{\mu}\hat{\nu}}$, summation over the indices $A, B= 1,2$ is implied, 
and $\epsilon_{AB}$ is antisymmetric with $\epsilon_{12}=+1$. It gives the 4D equation of motion
\begin{eqnarray}
\label{gaugino}
\Gamma^{\hat{\mu}}D_{\hat{\mu}}\Lambda_{A}+\frac{1}{2}\hat{G}\epsilon_{AB}\Lambda_{B}=0~.
\end{eqnarray}
Applying the partial wave expansion \eqref{Expansion:f3} to the two Majorana gaugini $\Lambda_A$ we find
\begin{eqnarray}
\label{Expansion:flr}
&&\Lambda^{(L/R)}_{A}=\frac{1}{2}(1\pm\Gamma_{5})\Lambda_{A}=\frac{1}{2}\left(\lambda_{A+}\pm\gamma_{A}\lambda_{A-}\right)\otimes\eta+\frac{1}{2}\left(\lambda_{A-}\pm\gamma_{A}\lambda_{A+}\right)\otimes\gamma_{S}\eta~.
\end{eqnarray}
for their left- and right-handed components. The indices $(\sigma lm)$ on the 2D fields $\lambda_{A\pm}$ and the spinor harmonics $\eta$ are suppressed for brevity. Inserting the expansion in spinor partial waves \eqref{Expansion:flr} into the 4D equations of motion \eqref{gaugino} 
projected on to the right helicity by the operator $\frac{1}{2}\left(1-\Gamma_5\right)$  
we get
\begin{eqnarray}
 &&0=\left[\gamma^{\mu}D_{\mu}\left(\lambda_{A-}+\gamma_{A}\lambda_{A+}\right)+i(l+1)\left(\lambda_{A+}+\gamma_{A}\lambda_{A-}\right)-\frac{1}{2}e^{i\frac{\pi}{4}}\epsilon_{AB}\left(\lambda_{B-}-\gamma_{A}\lambda_{B+}\right)\right]\otimes\eta\nonumber\\
&&+\left[\gamma^{\mu}D_{\mu}\left(\lambda_{A+}+\gamma_{A}\lambda_{A-}\right)-i(l+1)\left(\lambda_{A-}+\gamma_{A}\lambda_{A+}\right)-\frac{1}{2}e^{i\frac{\pi}{4}}\epsilon_{AB}\left(\lambda_{B+}-\gamma_{A}\lambda_{B-}\right)\right]\otimes\gamma_{S}\eta~.
\end{eqnarray}
Orthogonality of spinor harmonics then give us the 2D equation of motion 
\begin{eqnarray}
\label{eqn:gaugino:L:hat}
&&\gamma^{\mu}D_{\mu}\hat{\lambda}^{(L)}_{A}-(l+1)\hat{\lambda}^{(L)}_{A}-\frac{1}{2}e^{i\frac{\pi}{4}}\epsilon_{AB}\hat{\lambda}^{(R)}_{B}=0~,\\
\label{eqn:gaugino:L:tilde}
&&\gamma^{\mu}D_{\mu}\widetilde{\lambda}^{(L)}_{A}+(l+1)\widetilde{\lambda}^{(L)}_{A}-\frac{1}{2}e^{i\frac{\pi}{4}}\epsilon_{AB}\widetilde{\lambda}^{(R)}_{B}=0~,
\end{eqnarray}
for every spinor harmonic index $(\sigma lm)$. Here $\hat{\lambda}^{(L/R)}_{A}$ and $\widetilde{\lambda}^{(L/R)}_{A}$ are defined by
\begin{eqnarray}
\label{eqn:hattilde}
&&\hat{\lambda}^{(L/R)}_{A}\equiv\left(\lambda_{A+}\pm\gamma_{A}\lambda_{A-}\right)+i\left(\lambda_{A-}\pm\gamma_{A}\lambda_{A+}\right)~,\\
&&\widetilde{\lambda}^{(L/R)}_{A}\equiv\left(\lambda_{A+}\pm\gamma_{A}\lambda_{A-}\right)-i\left(\lambda_{A-}\pm\gamma_{A}\lambda_{A+}\right)~.
\end{eqnarray}
Similarly acting with the left projection operator $\frac{1}{2}\left(1+\Gamma_5\right)$ on the 4D equations of motion \eqref{gaugino}, we find the 2D wave equations that are conjugate of (\ref{eqn:gaugino:L:hat}, \ref{eqn:gaugino:L:tilde}):
\begin{eqnarray}
\label{eqn:gaugino:R:tilde}
&&\gamma^{\mu}D_{\mu}\widetilde{\lambda}^{(R)}_{B}+(l+1)\widetilde{\lambda}^{(R)}_{B}+\frac{1}{2}e^{-i\frac{\pi}{4}}\epsilon_{BA}\widetilde{\lambda}^{(L)}_{A}=0~,\\
\label{eqn:gaugino:R:hat}
&&\gamma^{\mu}D_{\mu}\hat{\lambda}^{(R)}_{B}-(l+1)\hat{\lambda}^{(R)}_{B}+\frac{1}{2}e^{-i\frac{\pi}{4}}\epsilon_{BA}\hat{\lambda}^{(L)}_{A}=0~.
\end{eqnarray}
Combining \eqref{eqn:gaugino:L:hat} and \eqref{eqn:gaugino:R:hat}, as well as \eqref{eqn:gaugino:L:tilde} and \eqref{eqn:gaugino:R:tilde} with $A=1,B=2$, we get
\begin{eqnarray}
\label{eqn:gaufi}
&&\left(\gamma^{\mu}D_{\mu}-(l+1)\right)\begin{pmatrix}\hat{\lambda}^{(L)}_{1} \\ \hat{\lambda}^{(R)}_{2}\end{pmatrix}=\frac{1}{2}\begin{pmatrix}0 & e^{i\frac{\pi}{4}} \\ e^{-i\frac{\pi}{4}} & 0\end{pmatrix}\begin{pmatrix}\hat{\lambda}^{(L)}_{1} \\ \hat{\lambda}^{(R)}_{2}\end{pmatrix}~.\\
&&\left(\gamma^{\mu}D_{\mu}+(l+1)\right)\begin{pmatrix}\widetilde{\lambda}^{(L)}_{1} \\ \widetilde{\lambda}^{(R)}_{2}\end{pmatrix}=\frac{1}{2}\begin{pmatrix}0 & e^{i\frac{\pi}{4}} \\ e^{-i\frac{\pi}{4}} & 0\end{pmatrix}\begin{pmatrix}\widetilde{\lambda}^{(L)}_{1} \\ \widetilde{\lambda}^{(R)}_{2}\end{pmatrix}~.
\end{eqnarray}
We are giving these results in full gory detail because the phases $e^{\pm i\frac{\pi}{4}}$ in the final result are physical consequences of the interplay between electric and magnetic fields which can be technically challenging to account for.  

The matrices on the right hand side of \eqref{eqn:gaufi} have eigenvalues $\pm\frac{1}{2}$. Therefore, the eigenvalues of the Dirac operator 
$\gamma^{\mu}D_{\mu}$ give the four AdS$_2$ spinor masses
\begin{eqnarray}
m=\pm (l+\frac{1}{2}) ~,\ \pm (l+\frac{3}{2})~. 
\end{eqnarray}
The sign of the fermion mass is formal and has no physical meaning. The conformal weight of the 1D conformal operator dual to an AdS$_2$ spinor is given by the relation $h_{\text{spinor}}= |m|+\frac{1}{2}$, so we find $h=l+1,l+2$, each with multiplicity $2$. 

The harmonic expansion for spinor fields has degeneracy $2(l+1)$, while the irreducible representation of $SU(2)$ labeled by the angular quantum number $j$ has $(2j+1)$ states. We therefore identify $j=l+\frac{1}{2}$ for spinors. This gives our final result for the spectrum of the gaugino block
\begin{eqnarray}
h= 2 \times (j+\frac{1}{2})~,\ ~2 \times (j+\frac{3}{2})~,
\end{eqnarray}
where ``$2\times$" denotes multiplicity 2, not the normal multiplication. This result is valid for all $j\geq \frac{1}{2}$.

\subsection{Bulk Modes of Gravitino Block}
\label{section:bulk:gravitino} 
  
The gravitino block has the 4D Lagrangian \cite{Castro:2018hsc}
\begin{eqnarray}\label{eqn:Lagrangoan:gravitino block}
e^{-1}\delta^{2}\mathcal{L}_{\text{gravitino}}&=&-\bar{\Psi}_{A\hat{\mu}}\Gamma^{\hat{\mu}\hat{\nu}\hat{\rho}}D_{\hat{\nu}}\Psi_{A\hat{\rho}}-2\bar{\Lambda}_{A}\Gamma^{\hat{\mu}}D_{\hat{\mu}}\Lambda_{A}-\frac{1}{2}\epsilon_{AB}\bar{\Psi}_{A\hat{\mu}}\left(G^{\hat{\mu}\hat{\nu}}+\Gamma_{5}\widetilde{G}^{\hat{\mu}\hat{\nu}}\right)\Psi_{B\hat{\nu}}\cr
&&-\frac{\sqrt{3}}{2}\left(\bar{\Psi}_{A\hat{\mu}}\hat{G}\Gamma^{\hat{\mu}}\Lambda_{A}+\bar{\Lambda}_{A}\Gamma^{\hat{\mu}}\hat{G}\Psi_{A\hat{\mu}}\right)+2\epsilon_{AB}\bar{\Lambda}_{A}\hat{G}\Lambda_{B}~,
\end{eqnarray}
where $\widetilde{G}^{\hat{\mu}\hat{\nu}}\equiv-\frac{i}{2}\epsilon^{\hat{\mu}\hat{\nu}\hat{\rho}\hat{\sigma}}G_{\hat{\rho}\hat{\sigma}}$. It gives the 4D equations of motion 
\begin{eqnarray}
\label{EoM:Psi}&&\Gamma^{\hat{\mu}\hat{\nu}\hat{\rho}}D_{\hat{\nu}}\Psi_{A\hat{\rho}}+\frac{1}{2}\left(G^{\hat{\mu}\hat{\nu}}+\Gamma_{5}\widetilde{G}^{\hat{\mu}\hat{\nu}}\right)\epsilon_{AB}\Psi_{B\hat{\nu}}+\frac{\sqrt{3}}{2}\hat{G}\Gamma^{\hat{\mu}}\Lambda_{A}=0~,\\
\label{EoM:Lambda}&&2\Gamma^{\hat{\mu}}D_{\hat{\mu}}\Lambda_{A}-2\epsilon_{AB}\hat{G}\Lambda_{B}+\frac{\sqrt{3}}{2}\Gamma^{\hat{\mu}}\hat{G}{\Psi}_{A\hat{\mu}}=0~.
\end{eqnarray}
In the following we work out the corresponding 2D equations of motion. 

\bigskip
\noindent
{\bf \emph{Gravitino block: gravitino equation} }

We first act with the right projection operator $\frac{1}{2}\left(1-\Gamma_5\right)$ on the 4D equations of motion for gravitini (\ref{EoM:Psi}) and then insert partial wave expansions in spinor harmonics (\ref{Expansion:f3}-\ref{Expansion:f1}). The $S^2$ components $\hat{\mu}=\alpha$ of the equations  become
%
\begin{eqnarray}
&&0=\left[\frac{1}{2}i(l+1)\gamma_{\mu}\left(\psi^{\mu}_{A-}+\gamma_{A}\psi^{\mu}_{A+}\right)+\gamma_{\mu\nu}D^{\mu}\left(\psi^{\nu}_{A+}+\gamma_{A}\psi^{\nu}_{A-}\right)+\frac{\sqrt{3}}{2}ie^{-i\frac{\pi}{4}}\left(\lambda_{A-}+\gamma_{A}\lambda_{A+}\right)\right]\otimes\gamma^{\alpha}\eta\nonumber\\
&&+\left[-\frac{1}{2}i(l+1)\gamma_{\mu}\left(\psi^{\mu}_{A+}+\gamma_{A}\psi^{\mu}_{A-}\right)+\gamma_{\mu\nu}D^{\mu}\left(\psi^{\nu}_{A-}+\gamma_{A}\psi^{\nu}_{A+}\right)+\frac{\sqrt{3}}{2}ie^{-i\frac{\pi}{4}}\left(\lambda_{A+}+\gamma_{A}\lambda_{A-}\right)\right]\otimes\gamma^{\alpha}\gamma_{S}\eta\nonumber\\
&&+\left[-\gamma_{\mu}\left(\psi^{\mu}_{A-}+\gamma_{A}\psi^{\mu}_{A+}\right)+\gamma^{\mu}D_{\mu}\left(\psi_{A-}+\gamma_{A}\psi_{A+}\right)-\frac{1}{2}ie^{i\frac{\pi}{4}}\epsilon_{AB}\left(\psi_{B-}-\gamma_{A}\psi_{B+}\right)\right]\otimes D^{(\alpha)}\eta\nonumber\\
&&+\left[-\gamma_{\mu}\left(\psi^{\mu}_{A+}+\gamma_{A}\psi^{\mu}_{A-}\right)+\gamma^{\mu}D_{\mu}\left(\psi_{A+}+\gamma_{A}\psi_{A-}\right)-\frac{1}{2}ie^{i\frac{\pi}{4}}\epsilon_{AB}\left(\psi_{B+}-\gamma_{A}\psi_{B-}\right)\right]\otimes D^{(\alpha)}\gamma_{S}\eta~.
\end{eqnarray}
%
The AdS$_2$ components $\hat{\mu}=\mu$ of the equations similarly give
\begin{eqnarray}
&&0=\left[-i(l+1)\gamma_{\mu\nu}\left(\psi^{\nu}_{A+}+\gamma_{A}\psi^{\nu}_{A-}\right)+\frac{1}{2}((l+1)^{2}-1)\gamma_{\mu}\left(\psi_{A-}+\gamma_{A}\psi_{A+}\right)\right.\nonumber\\ &&\left.-\frac{\sqrt{3}}{2}ie^{i\frac{\pi}{4}}\gamma_{\mu}\left(\lambda_{A+}+\gamma_{A}\lambda_{A-}\right)+\frac{1}{2}e^{-i\frac{\pi}{4}}\gamma_{\mu\nu}\epsilon_{AB}\left(\psi^{\nu}_{B-}-\gamma_{A}\psi^{\nu}_{B+}\right)\right]\otimes\eta\nonumber\\
&&+\left[i(l+1)\gamma_{\mu\nu}\left(\psi^{\nu}_{A-}+\gamma_{A}\psi^{\nu}_{A+}\right)+\frac{1}{2}((l+1)^{2}-1)\gamma_{\mu}\left(\psi_{A+}+\gamma_{A}\psi_{A-}\right)\right.\nonumber\\ &&\left. -\frac{\sqrt{3}}{2}ie^{i\frac{\pi}{4}}\gamma_{\mu}\left(\lambda_{A-}+\gamma_{A}\lambda_{A+}\right)
+\frac{1}{2}e^{-i\frac{\pi}{4}}\gamma_{\mu\nu}\epsilon_{AB}\left(\psi^{\nu}_{B+}-\gamma_{A}\psi^{\nu}_{B-}\right)\right]\otimes\gamma_{S}\eta~.
\end{eqnarray}
Again, we suppress the indices $(\sigma lm)$ of the spinor harmonics. Orthogonality of the spinor harmonics mean each square bracket vanishes by itself. This gives four towers of equations from the $S^2$ but only two from the AdS$_2$ because the $\hat{\mu}=\alpha$ index incorporates spin-$\frac{3}{2}$ components on $S^2$ while the $\hat{\mu}=\mu$ index only includes spin-$\frac{1}{2}$ components. To present the equations 
we define
\begin{eqnarray}
&&\hat{\psi}^{(L/R)}_{A\mu}\equiv\left(\psi_{A\mu+}\pm\gamma_{A}\psi_{A\mu-}\right)+i\left(\psi_{A\mu-}\pm\gamma_{A}\psi_{A\mu+}\right)~,\\
&&\widetilde{\psi}^{(L/R)}_{A\mu}\equiv\left(\psi_{A\mu+}\pm\gamma_{A}\psi_{A\mu-}\right)-i\left(\psi_{A\mu-}\pm\gamma_{A}\psi_{A\mu+}\right)~,
\end{eqnarray}
in analogy with the variables $\hat{\lambda}^{(L/R)}_{A}$ and $\widetilde{\lambda}^{(L/R)}_{A}$ introduced for the gaugino block in \eqref{eqn:hattilde}.
This gives the three coupled equations 
\begin{eqnarray}\label{EoM:gravitino:LR1}
&&\gamma^{\mu\nu}D_{\mu}\widetilde{\psi}^{(L)}_{A\nu}-\frac{1}{2}(l+1)\gamma^{\mu}\widetilde{\psi}^{(L)}_{A\mu}+\frac{\sqrt{3}}{2}e^{-i\frac{\pi}{4}}\hat{\lambda}^{(L)}_{A}=0~,\\\label{EoM:gravitino:LR2}
&&-\gamma^{\mu}\widetilde{\psi}^{(L)}_{A\mu}+\gamma^{\mu}D_{\mu}\widetilde{\psi}^{(L)}_{A}-\frac{1}{2}ie^{i\frac{\pi}{4}}\epsilon_{AB}\widetilde{\psi}^{(R)}_{B}=0~,\\\label{EoM:gravitino:LR3}
&&(l+1)\widetilde{\psi}^{(L)}_{A\rho}-\frac{1}{2}((l+1)^{2}-1)\gamma_{\rho}\widetilde{\psi}^{(L)}_{A}+\frac{\sqrt{3}}{2}e^{-i\frac{\pi}{4}}\hat{\lambda}^{(L)}_{A}-\frac{1}{2}e^{-i\frac{\pi}{4}}\epsilon_{AB}\widetilde{\psi}^{(R)}_{B\rho}=0~,
\end{eqnarray}
as well as the three coupled equations
\begin{eqnarray}\label{EoM:gravitino:LR1:conjugate}
&&\gamma^{\mu\nu}D_{\mu}\hat{\psi}^{(L)}_{A\nu}+\frac{1}{2}(l+1)\gamma^{\mu}\hat{\psi}^{(L)}_{A\mu}-\frac{\sqrt{3}}{2}e^{-i\frac{\pi}{4}}\widetilde{\lambda}^{(L)}_{A}=0~,\\\label{EoM:gravitino:LR2:conjugate}
&&-\gamma^{\mu}\hat{\psi}^{(L)}_{A\mu}+\gamma^{\mu}D_{\mu}\hat{\psi}^{(L)}_{A}-\frac{1}{2}ie^{i\frac{\pi}{4}}\epsilon_{AB}\hat{\psi}^{(R)}_{B}=0~,\\\label{EoM:gravitino:LR3:conjugate}
&&-(l+1)\hat{\psi}^{(L)}_{A\rho}-\frac{1}{2}((l+1)^{2}-1)\gamma_{\rho}\hat{\psi}^{(L)}_{A}-\frac{\sqrt{3}}{2}e^{-i\frac{\pi}{4}}\widetilde{\lambda}^{(L)}_{A}-\frac{1}{2}e^{-i\frac{\pi}{4}}\epsilon_{AB}\hat{\psi}^{(R)}_{B\rho}=0~.
\end{eqnarray}
Similarly, starting out by acting with the left projection operator $\frac{1}{2}\left(1+\Gamma_5\right)$ on the 4D equations of motion we find the complex conjugate of the preceding six equations.

\bigskip
\noindent
{\bf \emph{Gravitino block: gaugino equation} }
    
Acting with the right projection operator $\frac{1}{2}\left(1-\Gamma_5\right)$ on the 4D equations of motion for gaugini (\ref{EoM:Lambda}) and expanding it in partial waves, we get
\begin{eqnarray}
 &&\left[\gamma^{\mu}D_{\mu}\left(\lambda_{A-}+\gamma_{A}\lambda_{A+}\right)+i(l+1)\left(\lambda_{A+}+\gamma_{A}\lambda_{A-}\right)+ie^{-i\frac{\pi}{4}}\epsilon_{AB}\left(\lambda_{B-}-\gamma_{A}\lambda_{B+}\right)\right.\nonumber\\
&&\left.-\frac{\sqrt{3}}{4}ie^{i\frac{\pi}{4}}\gamma_{\mu}\left(\psi^{\mu}_{A+}+\gamma_{A}\psi^{\mu}_{A-}\right)\right]\otimes\eta+\nonumber\\
&&\left[\gamma^{\mu}D_{\mu}\left(\lambda_{A+}+\gamma_{A}\lambda_{A-}\right)-i(l+1)\left(\lambda_{A-}+\gamma_{A}\lambda_{A+}\right)+ie^{-i\frac{\pi}{4}}\epsilon_{AB}\left(\lambda_{B+}-\gamma_{A}\lambda_{B-}\right)\right.\nonumber\\
&&\left.-\frac{\sqrt{3}}{4}ie^{i\frac{\pi}{4}}\gamma_{\mu}\left(\psi^{\mu}_{A-}+\gamma_{A}\psi^{\mu}_{A+}\right)\right]\otimes\gamma_{S}\eta=0~.
\end{eqnarray}
Again, orthogonality implies that each square bracket vanishes by itself. After introduction of the variables \eqref{eqn:hattilde} the 2D equation of motion become
\begin{eqnarray}\label{EoM:gravitino:gaugino:LR}
&&\gamma^{\mu}D_{\mu}\hat{\lambda}^{(L)}_{A}-(l+1)\hat{\lambda}^{(L)}_{A}+ie^{-i\frac{\pi}{4}}\epsilon_{AB}\hat{\lambda}^{(R)}_{B}+\frac{\sqrt{3}}{4}e^{i\frac{\pi}{4}}\gamma^{\mu}\widetilde{\psi}^{(L)}_{A\mu}=0~,\\
\label{EoM:gravitino:gaugino:LR:conjugate}
&&\gamma^{\mu}D_{\mu}\widetilde{\lambda}^{(L)}_{A}+(l+1)\widetilde{\lambda}^{(L)}_{A}+ie^{-i\frac{\pi}{4}}\epsilon_{AB}\widetilde{\lambda}^{(R)}_{B}-\frac{\sqrt{3}}{4}e^{i\frac{\pi}{4}}\gamma^{\mu}\hat{\psi}^{(L)}_{A\mu}=0~.
\end{eqnarray}
 Similarly, starting out by acting with the left projection operator $\frac{1}{2}\left(1+\Gamma_5\right)$ on the 4D equations of motion we find the complex conjugate of these two equations.

We next proceed to solve the 2D equations of motion and compute the mass spectrum of gravitino block. 
We first discuss modes with $l\geq1$ modes and then deal with the special case $l=0$. 

\bigskip
\noindent
{\bf \emph{Gravitino block: $l\geq1$ modes} }
    
We begin by considering (\ref{EoM:gravitino:LR1}, \ref{EoM:gravitino:LR2}, \ref{EoM:gravitino:LR3}, \ref{EoM:gravitino:gaugino:LR}) and the equations conjugate to (\ref{EoM:gravitino:LR1:conjugate}, \ref{EoM:gravitino:LR2:conjugate}, \ref{EoM:gravitino:LR3:conjugate},\ref{EoM:gravitino:gaugino:LR:conjugate}) since only $\widetilde{\psi}^{(L/R)}_{A\mu}$, $\widetilde{\psi}^{(L/R)}_{A}$ and $\hat{\lambda}^{(L/R)}_{A}$ are involved.in this system.
    
Inspection of \eqref{EoM:gravitino:LR3} with $A=1$ and the conjugate of \eqref{EoM:gravitino:LR3:conjugate} with $A=2$ shows that the 2D gravitino $\widetilde{\psi}^{(L/R)}_{A\mu}$ is not an independent field. It can be expressed by $\widetilde{\psi}^{(L/R)}_{A}$ and $\hat{\lambda}^{(L/R)}_{A}$ as
\begin{eqnarray}
\label{eqn:psimu}
&&\begin{pmatrix}\widetilde{\psi}^{(L)}_{1\mu}\\ \widetilde{\psi}^{(R)}_{2\mu}\end{pmatrix}=\begin{pmatrix}\frac{4(l+1)}{4(l+1)^{2}-1} & \frac{2e^{-i\frac{\pi}{4}}}{4(l+1)^{2}-1} \\ \frac{2e^{i\frac{\pi}{4}}}{4(l+1)^{2}-1} & \frac{4(l+1)}{4(l+1)^{2}-1}\end{pmatrix}\gamma_{\mu}\left[\frac{1}{2}((l+1)^{2}-1)\begin{pmatrix}\widetilde{\psi}^{(L)}_{1} \\ \widetilde{\psi}^{(R)}_{2}\end{pmatrix}-\frac{\sqrt{3}}{2}\begin{pmatrix} e^{-i\frac{\pi}{4}}\hat{\lambda}^{(L)}_{1} \\ e^{i\frac{\pi}{4}}\hat{\lambda}^{(R)}_{2} \end{pmatrix} \right]~.
\end{eqnarray}
Inserting this into (\ref{EoM:gravitino:LR2}, \ref{EoM:gravitino:gaugino:LR}) with $A=1$ and the conjugates 
of (\ref{EoM:gravitino:LR2:conjugate}, \ref{EoM:gravitino:gaugino:LR:conjugate}) with $A=2$ we find
\begin{small}
\begin{eqnarray}
\label{EoM:Gravitino:Eigen}
&&\gamma^{\mu}D_{\mu}\begin{pmatrix}\hat{\lambda}^{(L)}_{1} \\ \hat{\lambda}^{(R)}_{2} \\ \widetilde{\psi}^{(L)}_{1}\\ \widetilde{\psi}^{(R)}_{2}\end{pmatrix}=\begin{pmatrix}l+1 & -e^{i\frac{\pi}{4}} &  0 & 0 \\ -e^{-i\frac{\pi}{4}} & l+1 & 0 & 0 \\ 0 & 0 & 0 & -\frac{1}{2}e^{-i\frac{\pi}{4}}\\ 0 & 0 &  -\frac{1}{2}e^{i\frac{\pi}{4}} & 0\end{pmatrix}\begin{pmatrix}\hat{\lambda}^{(L)}_{1} \\ \hat{\lambda}^{(R)}_{2} \\ \widetilde{\psi}^{(L)}_{1}\\ \widetilde{\psi}^{(R)}_{2}\end{pmatrix}+\gamma^{\mu}\begin{pmatrix}-\frac{\sqrt{3}}{4}e^{i\frac{\pi}{4}}\widetilde{\psi}^{(L)}_{1\mu} \\ -\frac{\sqrt{3}}{4}e^{-i\frac{\pi}{4}}\widetilde{\psi}^{(R)}_{2\mu} \\ \widetilde{\psi}^{(L)}_{1\mu} \\ \widetilde{\psi}^{(R)}_{2\mu} \end{pmatrix}\\
&&=\begin{pmatrix}(l+1)+\frac{3(l+1)}{4(l+1)^{2}-1} & \left(\frac{3(l+1)}{8(l+1)^{2}-2}-1\right)e^{i\frac{\pi}{4}} &  -\frac{\sqrt{3}(l+1)((l+1)^{2}-1)}{4(l+1)^{2}-1}e^{i\frac{\pi}{4}} & -\frac{\sqrt{3}((l+1)^{2}-1)}{8(l+1)^{2}-2}
 \\ \left(\frac{3(l+1)}{8(l+1)^{2}-2}-1\right)e^{-i\frac{\pi}{4}} & (l+1)+\frac{3(l+1)}{4(l+1)^{2}-1}  & -\frac{\sqrt{3}((l+1)^{2}-1)}{8(l+1)^{2}-2} & -\frac{\sqrt{3}(l+1)((l+1)^{2}-1)}{4(l+1)^{2}-1}e^{-i\frac{\pi}{4}}
\\ \frac{4\sqrt{3}(l+1)}{4(l+1)^{2}-1}e^{-i\frac{\pi}{4}} & -\frac{2\sqrt{3}}{4(l+1)^{2}-1} & (l+1)-\frac{3(l+1)}{4(l+1)^{2}-1}  & -\frac{3}{8(l+1)^{2}-2}e^{-i\frac{\pi}{4}}
 \\ -\frac{2\sqrt{3}}{4(l+1)^{2}-1} & \frac{4\sqrt{3}(l+1)}{4(l+1)^{2}-1}e^{i\frac{\pi}{4}} &  -\frac{3}{8(l+1)^{2}-2}e^{i\frac{\pi}{4}} & (l+1)-\frac{3(l+1)}{4(l+1)^{2}-1}  \end{pmatrix}
\begin{pmatrix}\hat{\lambda}^{(L)}_{1} \\ \hat{\lambda}^{(R)}_{2} \\ \widetilde{\psi}^{(L)}_{1}\\ \widetilde{\psi}^{(R)}_{2}\end{pmatrix}~.\nonumber
\end{eqnarray}
\end{small}
The matrix on the right hand side appears very complicated but, remarkably, it has simple eigenvalues: $l-\frac{1}{2}$, $l+\frac{1}{2}$, $l+\frac{3}{2}$, and $l+\frac{5}{2}$. 

Similarly, taking equations (\ref{EoM:gravitino:LR1:conjugate}, \ref{EoM:gravitino:LR2:conjugate}, \ref{EoM:gravitino:LR3:conjugate}, \ref{EoM:gravitino:gaugino:LR:conjugate}) with $A=1$ and the conjugates of (\ref{EoM:gravitino:LR1}, \ref{EoM:gravitino:LR2}, \ref{EoM:gravitino:LR3}, \ref{EoM:gravitino:gaugino:LR}) with $A=2$ we find a matrix equation for $\widetilde{\lambda}^{(L)}_{1}, \widetilde{\lambda}^{(R)}_{2},\hat{\psi}^{(L)}_{1}, \hat{\psi}^{(R)}_{2}$. The matrix again has simple eigenvalues: $-\left(l-\frac{1}{2}\right)$, $-\left(l+\frac{1}{2}\right)$, $-\left(l+\frac{3}{2}\right)$, and $-\left(l+\frac{5}{2}\right)$. Thus the complete result for the eigenvalues of $\gamma^{\mu}D_{\mu}$ can be expressed as eight AdS$_2$ spinor masses   
\begin{eqnarray}
m= \pm (l-\frac{1}{2})~,\ \pm(l+\frac{1}{2})~,\ \pm(l+\frac{3}{2})~,\ \pm(l+\frac{5}{2})~. 
\end{eqnarray}
For spinors we use the relations $j_{\text{spinor}}=l+\frac{1}{2}$ for the $SU(2)$ quantum number and $h_{\text{spinor}}=|m|+\frac{1}{2}$ for the conformal weight and so our result for the spectrum of the gravitino block becomes
\begin{eqnarray}
\label{eqn:gravires}
h= 2 \times (j-\frac{1}{2})~,\ 2 \times (j+\frac{1}{2})~,\ 2 \times (j+\frac{3}{2}) ~,\ 2 \times (j+\frac{5}{2})~,
\end{eqnarray}
where ``$2\times$" denotes multiplicity 2, not the normal multiplication. This result is valid for $j\geq \frac{3}{2}$.

\bigskip
\noindent
{\bf \emph{Gravitino block: $l=0$ modes} }
    
The $l=0$ mode is special for gravitini because the helicity $\pm{3\over 2}$ components in the partial wave expansion \eqref{Expansion:f2} vanish identically. Therefore, the fields $\psi_{A\pm}$ are not defined for $l=0$. 

The manipulations giving \eqref{eqn:psimu} for the AdS$_2$ gravitini remain valid for $l=0$ and we see that, in this special case, the term involving the nonexistent $\psi_{A\pm}$ has vanishing coefficient. Therefore, all components of the 2D gravitini $\psi_{A\mu}$ are determined by the gaugini $\lambda_A$. Accordingly, the first equation in \eqref{EoM:Gravitino:Eigen} depends only on gaugini for $l=0$. The resulting equation of motion can be read off from the upper left $2\times 2$ submatrix of the second equation in \eqref{EoM:Gravitino:Eigen} by taking $l=0$: 
\begin{eqnarray}
&&\gamma^{\mu}D_{\mu}\begin{pmatrix}\hat{\lambda}^{(L)}_{1} \\ \hat{\lambda}^{(R)}_{2}\end{pmatrix}=\begin{pmatrix}2 & -\frac{1}{2}e^{i\frac{\pi}{4}}
\\ -\frac{1}{2}e^{-i\frac{\pi}{4}} & 2  \end{pmatrix}
\begin{pmatrix}\hat{\lambda}^{(L)}_{1} \\ \hat{\lambda}^{(R)}_{2}\end{pmatrix}~.
\end{eqnarray}
The matrix on the right hand side has eigenvalues $\frac{3}{2}$ and $\frac{5}{2}$ and the analogous equations for $\widetilde{\lambda}^{(L/R)}_{A}$ similarly give $-\frac{3}{2}$ and $-\frac{5}{2}$. This corresponds to two modes with conformal weight $h=2$ and another two with $h=3$. These modes 
extend the towers $h=2\times (j+\frac{3}{2}),\ 2\times (j+\frac{5}{2})$ in \eqref{eqn:gravires} so they apply for all $j\geq \frac{1}{2}$.

 \subsection{Boundary Modes}
 \label{section:bndy}
Boundary modes are harmonic modes on AdS$_2$ which are formally pure gauge but in fact physical because the gauge functions that generate them are non-normalizable. There are no boundary modes for the scalar block or gaugino block because they involve no gauge symmetries. Thus all boundary modes come from vector blocks, the KK block, and gravitino blocks. This subsection determines the boundary modes of these three types of blocks in turns. 

Since boundary modes are somewhat subtle we proceed with special care. In each case we add gauge fixing terms and compute the full off-shell spectrum, along with the appropriate ghosts. This requires some additional effort. On the other hand, since the gauge functions underlying boundary modes are harmonic, they generally do not couple to bulk modes so the relevant field content remains manageable. 

\bigskip
\noindent
{\bf \emph{Boundary modes in vector blocks} }
    
For boundary modes in vector blocks we add the gauge fixing term $\left(\nabla_{\hat{\mu}}a^{\hat{\mu}}\right)^2$ to the Lagrangian (\ref{EoM:block:v}), with hatted variables denoting 4D indices as in previous sections. We can consistently ignore the scalar field in the vector block because it couples to $\nabla_{\mu}b^{\mu}$ which vanishes in the boundary sector due to the harmonic condition. The effective Lagrangian of the boundary modes in the vector block becomes 
\begin{eqnarray}
\mathcal{L}^{\textrm{bndy}}_{\textrm{vector}}&=&b^{\mu}\left(\nabla^2_A+1-l(l+1)\right)b_{\mu}~.
\end{eqnarray}
Harmonic vector modes satisfy $\nabla^2_A+1$ so this is equivalent to a tower of nondynamical fields with $m^2=l(l+1)$, $l\geq0$ with degeneracy $2l+1$. This result is unsurprising because the residual gauge transformations underlying these modes satisfy the massless Klein-Gordon equation in 4D $\nabla^2_4 \Lambda=0$.

\bigskip
\noindent
{\bf \emph{Boundary modes in the KK block}}

In this sector we must consider gravity as well as the KK vector field. We add the gauge fixing term
\begin{eqnarray}
\mathcal{L}^{\textrm{g.f.}}_{\textrm{KK}}&=&-\left(\nabla^{\hat{\mu}}h_{\hat{\mu}\hat{\rho}}-\frac{1}{2}\nabla_{\hat{\rho}}h\right)\left(\nabla^{\hat{\nu}}h_{\hat{\nu}}^{\ \hat{\rho}}-\frac{1}{2}\nabla^{\hat{\rho}}h\right)-4\left(\nabla_{\hat{\mu}}a^{\hat{\mu}}\right)^2~,
\end{eqnarray}
to the 4D Lagrangian of the KK block. We then compute the corresponding 2D Lagrangian
\begin{eqnarray}
\label{eqn:bndyL}
&&\mathcal{L}^{\textrm{bndy}}_{\textrm{KK}}=H'^{\{\mu\nu\}}\left(\nabla^{2}_{x}+2-l(l+1)\right)H'_{\{\mu\nu\}}\\
&&+\begin{pmatrix}
b'_{\mu} & B'_{2\mu} & B'_{1\mu}
\end{pmatrix}
\begin{pmatrix}
\left(\nabla_{A}^{2}+1-l(l+1)\right)\delta^{\mu}_{\nu} & -\sqrt{2l(l+1)}\delta^{\mu}_{\nu} & -\sqrt{2l(l+1)}\epsilon^{\mu}_{\ \nu}\\
-\sqrt{2l(l+1)}\delta^{\mu}_{\nu} & \left(\nabla_{A}^{2}+1-l(l+1)\right)\delta^{\mu}_{\nu} & -2\epsilon^{\mu}_{\ \nu}\\
\sqrt{2l(l+1)}\epsilon^{\mu}_{\ \nu} & 2\epsilon^{\mu}_{\ \nu} & \left(\nabla_{A}^{2}+1-l(l+1)\right)\delta^{\mu}_{\nu}
\end{pmatrix}
\begin{pmatrix}
b'^{\nu} \\ B'^{\nu}_{2} \\ B'^{\nu}_{1}
\end{pmatrix}~,\nonumber
\end{eqnarray}
where we introduced conveniently normalized fields $H'_{\{\mu\nu\}}=\frac{1}{\sqrt{2}}H_{\{\mu\nu\}}$, $b'_{\mu}=2b_{\mu}$ and $B'^{\mu}_{1/2}=\sqrt{l(l+1)}B^{\mu}_{1/2}$. We consistently ignored scalar fields because their couplings to AdS$_2$ vectors all contain 
$\nabla_{\mu}b^{\mu}$ or $\nabla_{\mu}B^{\mu}_{1/2}$ which vanish for boundary modes.
    
We can diagonalize the mass matrix in $\mathcal{L}^{\textrm{bndy}}_{\textrm{KK}}$ above and so determine the eigenvalues of $\nabla^2_A$. The scalars that are equivalent to AdS$_2$ tensors and vectors have masses given by the eigenvalues of $\nabla^2_A+2$ and $\nabla^2_A+1$, respectively. The eigenvectors and the corresponding scalar masses become:
\begin{eqnarray}
m^2=l(l+1)&:&H'_{\{\mu\nu\}}~,\ l\geq0~,\\
\label{eqn:mixgauge}
m^2=l(l-1)&:&\sqrt{\frac{1+l}{1+2l}}b'_{\mu}-\sqrt{\frac{l}{1+2l}}\frac{1}{\sqrt{2}}\left(B'_{2\mu}+\epsilon_{\mu\nu}B'^{\nu}_{1}\right)~,\ l\geq0~,\\
m^2=(l+1)(l+2)&:&\sqrt{\frac{l}{1+2l}}b'_{\mu}+\sqrt{\frac{1+l}{1+2l}}\frac{1}{\sqrt{2}}\left(B'_{2\mu}+\epsilon_{\mu\nu}B'^{\nu}_{1}\right)~,\ l\geq1~,\\
m^2=l(l+1) + 2&:&\frac{1}{\sqrt{2}}\left(B'_{1\mu}+\epsilon_{\mu\nu}B'^{\nu}_{2}\right)~,\ l\geq1~.
\end{eqnarray}
The last three entries deserve some comments: the matrix in the second line of \eqref{eqn:bndyL} acts on AdS$_2$ vectors so it is $6\times 6$ and it allows $6$ eigenvectors. However, harmonic modes satisfy a duality condition so, in order to avoid overcounting we should take either the eigenvector or the dual eigenvector, not both. Since the formalism is off-shell there is some scheme-dependence to this choice. The analogous treatment of boundary modes for BPS black holes by Sen \cite{Sen:2011ba} includes all contributions and divide by two in the end. We elect instead to pick an orthogonal set that diagonalizes the off-diagonal terms in \eqref{eqn:bndyL} and cancels the ghost tower determined below. This choice seems more physical to us but the final on-shell results are at any rate independent of scheme. 

Tensors $H_{\{\mu\nu\}}$ have degeneracy three \cite{Sen:2011ba,Larsen:2014bqa}, therefore the boundary modes in the KK block are $3$ towers of $m^2=l(l+1)$ with $l\geq0$, $1$ tower of $m^2=l(l-1)$ with $l\geq0$, $1$ tower of $m^2=(l+1)(l+2)$ with $l\geq1$, and $1$ tower of $m^2=l(l+1)+2$ with $l\geq1$.

In the off-shell formalism that we apply for boundary modes we must consider also the contribution from the ghost that generates the diffeomorphism $\delta H_{\{\mu\nu\}}=\nabla_{\mu}\xi_{\nu}+\nabla_{\nu}\xi_{\mu}-g_{\mu\nu}\nabla^{\rho}\xi_{\rho}$. The ghost equation of motion follows by variation of the gauge condition under a diffeomorphism: 
\begin{eqnarray}
&&\delta\left(\nabla^{\hat{\mu}}h_{\hat{\mu}\hat{\rho}}-\frac{1}{2}\nabla_{\hat{\rho}}h\right)=0\quad\Rightarrow\quad \left(\nabla^2_A-1-l(l+1)\right)\xi_{\rho}=0~.
\end{eqnarray}
Since eigenvalues of $\nabla^2_A+1$ acting on a vector can be identified with the mass of the dual scalar we find that the boundary ghosts have $m^2=l(l+1)+2, l\geq0$ with degeneracy $2l+1$. These contributions effectively cancel one of the $6$ towers of KK boundary modes.
  
\bigskip
\noindent
{\bf \emph{Boundary modes in gravitino blocks}}
    
In the off-shell formalism that we apply to boundary modes we add the gauge fixing term $\frac{1}{2}\left(\bar{\Psi}_{A\hat{\mu}}\Gamma^{\hat{\mu}}\right)\Gamma^{\hat{\nu}}D_{\hat{\nu}}\left(\Gamma^{\hat{\rho}}{\Psi}_{A\hat{\rho}}\right)$ to the gravitino Lagrangian \eqref{eqn:Lagrangoan:gravitino block} and redefine the field as $\Phi_{A\hat{\mu}}={\Psi}_{A\hat{\mu}}-\frac{1}{2}\Gamma_{\hat{\mu}}\Gamma^{\hat{\rho}}{\Psi}_{A\hat{\rho}}$. The Lagrangian for the gravitini then becomes
\begin{eqnarray}
&&e^{-1}\delta^{2}\mathcal{L}^{\textrm{bndy}}_{\textrm{gravitino}}=-\bar{\Phi}_{A\hat{\mu}}H^{\hat{\mu}\hat{\nu}}_{AB}\Phi_{B\hat{\nu}}=-\bar{\Phi}_{A\hat{\mu}}\left[\Gamma^{\hat{\rho}}D_{\hat{\rho}}g^{\hat{\mu}\hat{\nu}}-\frac{1}{2}\epsilon_{AB}\left(G^{\hat{\mu}\hat{\nu}}+\Gamma_{5}\widetilde{G}^{\hat{\mu}\hat{\nu}}\right)\right]\Phi_{B\hat{\nu}}~,
\end{eqnarray}
after consistently ignoring the gaugini which do not couple to harmonic modes.
    
The square of the quadratic fluctuation operator $H^{\hat{\mu}\hat{\nu}}_{AB}$ is
\begin{eqnarray}
\Lambda^{\hat{\mu}\hat{\nu}}_{AB}=H^{\hat{\mu}\hat{\rho}\dagger}_{AC}H^{\ \hat{\nu}}_{\hat{\rho}\ CB}&=&-\left(\Gamma^{\hat{\rho}}D_{\hat{\rho}}\Gamma^{\hat{\sigma}}D_{\hat{\sigma}}+\frac{1}{4}\right)g^{\hat{\mu}\hat{\nu}}\delta_{AB}~.
\end{eqnarray}
Only the components of $\Phi^{\hat{\mu}}_{A}$ with $\hat{\mu}=\mu$ have boundary modes. Using the partial wave expansion $\Phi^{{\mu}}_{A}=\phi^{{\mu}}_{A+}\otimes\eta+\phi^{{\mu}}_{A-}\otimes\gamma_{S}\eta$ , the equation of motion $\Lambda^{{\mu}\hat{\nu}}_{AB}\Phi_{B\hat{\nu}}=0$ can be expanded as
\begin{eqnarray}
-\left(\gamma^{{\rho}}D_{{\rho}}\gamma^{{\sigma}}D_{{\sigma}}-(l+1)^{2}+\frac{1}{4}\right)\phi^{{\mu}}_{A+}\otimes\eta-\left(\gamma^{{\rho}}D_{{\rho}}\gamma^{{\sigma}}D_{{\sigma}}-(l+1)^{2}+\frac{1}{4}\right)\phi^{{\mu}}_{A-}\otimes\gamma_{S}\eta=0~. 
\end{eqnarray}
Thus we find that there are $4$ towers of gravitino boundary modes with identical mass squared $m^2=(l+1)^2-\frac{1}{4}$, each with the degeneracy $2l+2$.

\subsection{Summary of the Mass Spectrum}\label{section:spectrum summary}
As conclusion to this long section we summarize our results. 

\renewcommand{\arraystretch}{1.75}
\begin{table}[htbp]
\centering
\scalebox{1}[1]{
\begin{tabular}{|c|c|}
		\hline
		Blocks &	Bulk Modes Spectrum $(h,j)$	\\\hline\hline
		Scalar & $(k+1,k)$  \\\hline
		Gaugino &	$2(k+2,k+\frac{1}{2}) \qquad 2(k+1,k+\frac{1}{2})$  \\\hline
		Vector & $(k+2,k)\qquad (k+2,k+1)\qquad (k+1,k+1)$ \\\hline
		Gravitino &	$2(k+3,k+\frac{1}{2}) \qquad 2(k+2,k+\frac{1}{2}) \qquad 2(k+2,k+\frac{3}{2}) \qquad 2(k+1,k+\frac{3}{2})$  \\\hline
		KK & 	$(k+3,k) \qquad (k+3,k+1) \qquad (k+2,k+1) \qquad (k+2,k+2)\qquad (k+1,k+2)$\\
		\hline
\end{tabular}}
\caption{Mass spectrum of bulk modes in non-BPS blocks. The label $k=0,1,\ldots$}
\label{summary:h}
\end{table}
\renewcommand{\arraystretch}{1}

The mass spectrum $(h,j)$ for the bulk modes is given in table \ref{summary:h}. In all cases the conformal weight $h$ is related to an effective scalar mass as $m^2=h(h-1)$ for bosons or $m^2=(h-\frac{1}{2})^2$ for fermions. For scalar fields $m^2$ is the on-shell eigenvalue of $\nabla^2_A$. However, for vectors and tensors we identify $m^2$ as the eigenvalue of $\nabla^2_A+1$ and $\nabla^2_A+2$, respectively. This is justified by the action of the operators $\nabla^2_A+1$, $\nabla^2_A+2$ on vectors and tensors being equivalent to the action of $\nabla^2_A$ on the corresponding scalar field obtained by the appropriate dualization in AdS$_2$. Thus we can use the formula for the conformal weight \eqref{eqn:hform} for all bosons. 

A notable feature of the results recorded in table \ref{summary:h} is that the conformal weight $h$ is integral in all cases. Since $h$ is determined for each entry by solving the quadratic $m^2=h(h-1)$ or $m^2=(h-\frac{1}{2})^2$ this is a rather nontrivial result. It requires that all effective masses are such that the discriminant of the quadratic is a perfect square. This property would be expected if the spectrum was organized in supermultiplets but, in the present context, supersymmetry is entirely broken by the background. We will develop this point further in sections \ref{section:3D to 2D} and \ref{section:global symmetry}.

The angular momentum quantum number $j$ labels the irreducible representation of $SU(2)$. There are $(2j+1)$ states for each value of $j$. 
The values of $j$ are integral (half-integral) for bosons (fermions), as expected. It is interesting that, in contrast, the conformal weight $h$ is integral for both bosons and fermions. In our context states are not organized in supermultiplets so there is no general expectation that $h$ must be half-integral for fermions but the result seems surprising nonetheless. We will also develop this point further in section \ref{section:3D to 2D}.

The scalars in the vector block generally mix with the vector field. However, the vector field does not include a spherically symmetric mode so the $j=0$ sector has just one mode, an effective 2D scalar with $h=2$. A minimally coupled scalar would have $h=1$, as for the scalar block, so these scalars are non-minimally coupled even in the spherically symmetric sector. This is an aspect of the attractor mechanism which determines the horizon value of the scalars in the vector block as a function of the charges and therefore inhibits their fluctuations around the preferred attractor value. This is a nonBPS version of the mechanism familiar from BPS black holes where these fields are known as fixed scalars \cite{Callan:1996tv}. 

The analogous scalar mode in the $j=0$ sector of the KK block is also interesting. It has conformal weight $h=3$. Thus the coupling between the KK scalar and gravity is stronger than the analogous coupling between gauge fields and their scalar partners. This effect has no analogue on the BPS branch but the $h=3$ mode was previously identified for rotating black holes \cite{Castro:2018ffi}.

\renewcommand{\arraystretch}{1.25}
\begin{table}[htbp]
\centering
\begin{tabular}{|c|c|c|c|c|}
\hline
Blocks &	Masses of Boundary Modes & Degeneracy & Multiplicity	& Range\\\hline\hline
KK & $m^2=l(l+1)$ & $2l+1$ & 3 & $l\geq0$ \\\hline
KK & $m^2=(l-1)l$ & $2l+1$ & 1 & $l\geq0$ \\\hline
KK & $m^2=(l+1)(l+2)$ & $2l+1$ & 1 & $l\geq1$ \\\hline
KK & $m^2=l(l+1)+2$ & $2l+1$ & 1 & $l\geq1$ \\\hline
KK & $m^2=l(l+1)+2$ & $2l+1$ & $-1$ & $l\geq0$ \\\hline
Vector & $m^2=l(l+1)$ & $2l+1$ & 1 & $l\geq0$ \\\hline
Gravitino & $m^2=(l+1)^2-\frac{1}{4}$ & $2l+2$ & 4 & $l\geq0$ \\
\hline
\end{tabular}
\caption{Mass spectrum of boundary modes in non-BPS blocks. (Multiplicity $-1$ denotes contribution from ghosts.)}
\label{summary:h:bndy}
\end{table}
\renewcommand{\arraystretch}{1}
Boundary modes are more subtle since they are based on harmonic modes which have no bulk kinetic term. For these modes we worked out the full off-shell spectrum, to circumvent any ambiguity. The result is present in terms of ``masses" in table \ref{summary:h:bndy}. The mass indicates the departure from a true zero-mode so $m^2$ is the eigenvalue appropriate for computing functional determinants.


\section{Heat Kernels}\label{section:heat kernels}
In this section we use the mass spectrum determined in the previous section to compute the 4D heat kernel and the associated logarithmic corrections to black hole entropy. There are three distinct contributions:
 \begin{itemize}
 \item
\emph{Bulk modes}: The propagating degrees of freedom summarized in table \ref{summary:h}.
\item \emph{Boundary modes}: 
The global degrees of freedom due to the harmonic modes of $AdS_{2}$ vectors, gravitini, and tensors. They are summarized in table \ref{summary:h:bndy}.    	
\item \emph{Zero mode corrections}: 
On-shell boundary modes that were already counted as boundary modes need corrections to their counting weights.
\end{itemize}
%

\subsection{Heat Kernel Preliminaries}
The action for quadratic fluctuations around a background has the generic form
\begin{eqnarray}
\mathcal{S}=-\int d^4x\sqrt{-g}~\phi_{n}\Lambda^{n}_{m}\phi^{m}~,
\end{eqnarray}
where $\{\phi_n\}$ is a complete set of fields and $\Lambda^{n}_{m}$ is a matrix that encodes the action of quadratic fluctuations around the background. The heat kernel of the operator $\Lambda$ is then defined by
\begin{eqnarray}
K_{4}(s)=\textrm{Tr}~e^{-s\Lambda}=\sum_{i}e^{-s\lambda_i}~,
\label{eqn:K3def}
\end{eqnarray}
where $\{\lambda_{i}\}$ is the set of eigenvalues of $\Lambda$.

We denote the heat kernel of a massless field on AdS$_2$ by $K_{A}(s)$. The result for a mode with effective 2D mass $m^2$ 
is suppressed by an additional factor $e^{-m^2 s}$ so, upon summing over a complete tower of states with masses $m^2_j$ and $SU(2)$ quantum number $j$, we find 
\begin{eqnarray}
K_{4}(s)=K_{A}(s)K_{S}(s)=K_{A}(s)~\frac{1}{4\pi}\sum_{j} (2j+1)e^{-m_j^2 s}~,
\end{eqnarray}
for the 4D heat kernel on AdS$_2\times S^2$. The sum over the tower can be interpreted as a field on $S^2$ and so we divide it by the area $4\pi$ of the unit $S^2$ and denote it by $K_{S}(s)$. The masses are related to conformal weights as $m^2=h(h-1)$ for bosons and $m^2=(h-{1\over 2})^2=h(h-1) + {1\over 4}$ for fermions. 
    
The Laurent expansion of the heat kernel $K_{4}(s)$ around $s=0$ generally has poles of order $s^{-2}$ and $s^{-1}$, followed by a constant that we denote $K^{\textrm{const}}_4$. It is related to the central charge $a$ of the 4D conformal anomaly by 
\begin{eqnarray}
a=2\pi^2 K^{\textrm{const}}_4~.
\label{eqn:adef}
\end{eqnarray}
The other central charge $c$ is immaterial here because the AdS$_2\times S^2$ background is conformally flat. 

\subsection{Bulk Modes}
All bulk bosons in 2D are represented as scalars. A massless scalar on Euclidean AdS$_2$ has continuous eigenvalues $\lambda_{A}=p^2+\frac{1}{4},~p\in \mathbb{R}$ weighted by the Plancherel measure $\mu(p)=p\tanh(\pi p)$. It has heat kernel \cite{Camporesi:1994ga}: 
\begin{eqnarray}
K^{b}_{A}(s)&=&\frac{1}{2\pi}\int_{0}^{\infty}dp~ p\tanh{(\pi p)}\exp{\left[-s\left(p^2+\frac{1}{4}\right)\right]}\cr
&=&\frac{1}{4\pi s}\left(1-\frac{1}{3}s+\frac{1}{15}s^2+\mathcal{O}(s^3)\right)~.
\end{eqnarray}

For sums over towers of modes an essential benchmark is the heat kernel of a minimally coupled scalar on $S^2$. The standard result from introductory quantum mechanics is that the eigenvalues of $-\nabla_{S}^{2}$ are $l(l+1)$ with degeneracy $2l+1$ and range $l=0,1.\ldots$. This gives the heat kernel:
\begin{eqnarray}
K^{b}_{S}(s)=\frac{1}{4\pi}\sum^{\infty}_{k=0}(2k+1)e^{-k(k+1)s}=\frac{1}{4\pi s}\left(1+\frac{1}{3}s+\frac{1}{15}s^2+\mathcal{O}(s^3)\right)~.
\label{eqn:KS}
\end{eqnarray}

With these results the spectrum for bulk bosons given in table \ref{summary:h} yields the following heat kernels: 

\bigskip
\noindent
{\bf \emph{Scalar block} }

The scalar block is just a minimal scalar with spectrum $(h,j)=(k+1,k)$ for $k\geq0$. For bosons, we have $m^2=h(h-1)$ with degeneracy $2j+1$, therefore
\begin{eqnarray}
K^{\text{scalar}}_{4}&=&K^{b}_{A}K^{\text{scalar}}_{S}= K^{b}_{A}~ \frac{1}{4\pi}\sum_{k=0}(2k+1)e^{-k(k+1)s}\nonumber\\
&=&\frac{1}{16\pi^{2}s^{2}}\left(1+\frac{1}{45}s^2+\mathcal{O}(s^3)\right)~.
\end{eqnarray}
where we used \eqref{eqn:KS} for the sum over the tower. The constant term $K_4^{\textrm{const}}=\frac{1}{720\pi^2}$ corresponds to the conformal anomaly $a^{\text{scalar.bulk}}=\frac{1}{360}$ according to \eqref{eqn:adef}. This is the standard answer for a minimally coupled scalar (1 d.o.f.) \cite{Duff:1993wm,Vassilevich:2003xt}. 

\bigskip
\noindent
{\bf \emph{Vector block} }

 The spectrum of the vector block has $3$ towers: $(h,j)=(k+2,k)$, $(k+2,k+1)$, $(k+1,k+1)$ for $k\geq0$. Therefore we have
 \begin{eqnarray}
 K^{\text{vector}}_{4}&=&K^{b}_{A}K^{\text{vector}}_{S}\nonumber\\
 &=&\frac{K^{b}_{A}}{4\pi}\left(\sum_{k=0}(2k+1)e^{-(k+1)(k+2)s}+\sum_{k=1}(2k+1)\left(e^{-k(k+1)s}+e^{-(k-1)ks}\right)\right)\nonumber\\
 &=&K^{b}_{A}~ \frac{1}{4\pi}\left(3\sum_{k=0}(2k+1)e^{-k(k+1)s}\right)=3K^{\text{scalar}}_{4}~.
 \end{eqnarray}
Thus the vector block ($3$ d.o.f.) has the same heat kernel as $3$ minimally coupled scalars: $a^{\text{vector.bulk}}=\frac{1}{120}$.

\bigskip
\noindent
{\bf \emph{KK block} }
    
The spectrum of the KK block has $5$ towers: $(h,j)=(k+3,k)$, $(k+3,k+1)$, $(k+2,k+1)$, $(k+2,k+2)$, $(k+1,k+2)$ for $k\geq0$. Therefore we have 
\begin{eqnarray}
K^{\text{gravity}}_{4}&=&K^{b}_{A}K^{\text{gravity}}_{S}\nonumber\\
&=&\frac{K^{b}_{A}}{4\pi}\left(\sum_{k=0}(2k+1)e^{-(k+2)(k+3)s}+\sum_{k=1}(2k+1)\left(e^{-(k+1)(k+2)s}+e^{-k(k+1)s}\right)\right.\nonumber\\
&&\qquad\qquad\left.+\sum_{k=2}(2k+1)\left(e^{-(k-1)ks}+e^{-(k-2)(k-1)s}\right)\right)\nonumber\\
&=&K^{b}_{A}~\frac{1ß}{4\pi}\left(5\sum_{k=0}(2k+1)e^{-k(k+1)s}\right)=5K^{\text{scalar}}_{4}~.
\end{eqnarray}
Thus the KK block ($5$ d.o.f.) has the same heat kernel as $5$ minimally coupled scalars: $a^{\text{KK.bulk}}=\frac{1}{72}$.

\bigskip   
The heat kernel of a massless minimally coupled spinor (1 d.o.f.) on AdS$_2$ is given by \cite{Camporesi:1995fb}
\begin{eqnarray}
K^{{f}}_{A}(s)&=&-\frac{1}{2\pi}\int_{0}^{\infty}dp~p\coth{(\pi p)}\exp{\left(-sp^2\right)}\cr
&=&-\frac{1}{4\pi s}\left(1+\frac{1}{6}s-\frac{1}{60}s^2+\mathcal{O}(s^3)\right)~,
\end{eqnarray}
where the overall sign incorporates fermionic statistics. With this result as starting point, the spectrum for bulk fermions given in table \ref{summary:h} yields: 

\bigskip
\noindent
{\bf \emph{Gaugino block} }
    
The spectrum of the gaugino block has $4$ towers: two copies of $(h,j)=(k+2,k+\frac{1}{2})$, $(k+1,k+\frac{1}{2})$ for $k\geq0$. For fermions we have $m^{2}=(h-\frac{1}{2})^2=h(h-1)+\frac{1}{4}$ with degeneracy $2j+1$, therefore
\begin{eqnarray}
K^{\text{gaugino}}_{4}&=&K^{f}_{A}K^{\text{gaugino}}_{S}\nonumber\\
&=&K^{f}_{A}~\frac{2}{4\pi}e^{-\frac{1}{4}s}\left(\sum_{k=0}(2k+2)e^{-k(k+1)s}+\sum_{k=1}2ke^{-k(k+1)s}\right)\nonumber\\
&=&K^{f}_{A}~\frac{4}{4\pi}e^{-\frac{1}{4}s} \sum_{k=0}(2k+1)e^{-k(k+1)s} \nonumber\\
&=&-\frac{1}{4\pi^{2}s^{2}}\left(1+\frac{1}{4}s+\frac{17}{1440}s^2+\mathcal{O}(s^3)\right)~.
\end{eqnarray}
We used \eqref{eqn:KS} for the sum over the tower, as for bosons. The constant term $K^{\textrm{const}}_4=-\frac{17}{5760\pi^2}$
and \eqref{eqn:adef} give the conformal anomaly $a^{\text{gaugino.bulk}}=-\frac{17}{2880}$ for the gaugino block ($4$ d.o.f.).

\bigskip
\noindent
{\bf \emph{Gravitino block} }
    
The spectrum of the gravitino block has $8$ towers: two copies of $(h,j)=(k+3,k+\frac{1}{2})$, $(k+2,k+\frac{1}{2})$, $(k+2,k+\frac{3}{2})$, $(k+1,k+\frac{3}{2})$ for $k\geq0$. It gives the heat kernel
\begin{eqnarray}
K^{\text{gravitino}}_{4}&=&K^{f}_{A}K^{\text{gravitino}}_{S}\nonumber\\
&=&K^{f}_{A}~\frac{2}{4\pi}e^{-\frac{1}{4}s}\left(\sum_{k=0}(2k+4)e^{-k(k+1)s}+\sum_{k=1}(2k+2)e^{-k(k+1)s}\right.\nonumber\\
&&\qquad\qquad\qquad\left.+\sum_{k=1}2ke^{-k(k+1)s}+\sum_{k=2}(2k-2)e^{-k(k+1)s}\right)\nonumber\\
&=&K^{f}_{A}~\frac{8}{4\pi}e^{-\frac{1}{4}s} \sum_{k=0}(2k+1)e^{-k(k+1)s} =2K^{\text{gaugino}}_{4}~.
\end{eqnarray}
Thus the gravitino block ($8$ d.o.f.) has the same heat kernel as $2$ gaugino blocks: $a^{\text{gravitino.bulk}}=-\frac{17}{1440}$.   

\bigskip
It is interesting that in all cases the results are equivalent to free massless bosons or fermions with the appropriate number of degrees of freedom. This amounts to a delicate conspiracy between non-minimal couplings and ranges of partial wave towers. The origin of these simplifications is not clear to us. 
    
For ${\cal N}=8$ SUGRA, there are 1 KK block, 27 vector blocks, 42 minimally coupled scalars, 4 gravitino blocks, and 24 gaugino blocks. In this case the total contribution from the bulk modes becomes: 
\begin{eqnarray}
&&a^{\text{bulk}}_{\mathcal{N}=8}=(5+27\times3+42)\times\frac{1}{360}-(4\times2+24)\times\frac{17}{2880}=\frac{1}{6}~.
\end{eqnarray}

For $\mathcal{N}=4$ SUGRA with $n_V$ matter multiplets, there are 1 KK block, ($n_V+5$) vector blocks, ($5n_V-4$) minimally coupled scalars, 2 gravitino blocks, and $2n_V$ gaugino blocks, which give the bulk contribution $a^{\text{bulk}}_{\mathcal{N}=4}=\frac{n+2}{96}$.
\subsection{Boundary Modes}
As discussed in subsection \ref{section:bndy}, boundary modes are due to the harmonic modes on AdS$_{2}$ of vectors, gravitini, and tensors. The scalar and gaugino blocks do not have boundary modes. These modes are constant on the AdS$_2$ space with (renormalized) volume $2\pi$. Therefore, the heat kernel for a single boundary mode is given by
\begin{eqnarray}
K^{\textrm{zero}}_{A}(s)=\pm\frac{1}{2\pi}~,
\end{eqnarray}
where $\pm$ is for bosons/fermions. The contributions to the heat kernel from the entire towers of boundary modes are then computed as follows. 

\bigskip
\noindent
{\bf \emph{Vector block} }

The spectrum of boundary modes for the vector block given in table \ref{summary:h:bndy} is $m_{l}^{2}=l(l+1)$ with integral $l\geq0$. This is equivalent to a single scalar field on the $S^2$. Their contribution to the heat kernel become
\begin{eqnarray}
K^{\text{vector.bndy}}_{4}&=&K^{\text{zero}}_{A}K_{S}\nonumber\\
&=&\frac{1}{2\pi}\frac{1}{4\pi}\sum_{k=0}(2k+1)e^{-k(k+1)s}\nonumber\\
&=&\frac{1}{8\pi^{2}s}\left(1+\frac{1}{3}s+\mathcal{O}(s^2)\right)~,
\end{eqnarray}
where we used the sum \eqref{eqn:KS}.

According to \eqref{eqn:adef} the constant term in this expression gives conformal anomaly $a^{\text{vector.bndy}}=\frac{1}{12}$, so, 
adding the bulk contribution of a single vector block $a^{\text{vector.bulk}}=\frac{1}{120}$ from table \ref{summary:h}, our explicit sum over modes gives $a^{\text{vector.bulk+bndy}}=\frac{11}{120}$. This agrees with the result found in \cite{Castro:2018hsc} using a very different method.

\bigskip
\noindent 
{\bf \emph{KK block} }

The boundary modes listed for the KK block in table \ref{summary:h:bndy} comprise $6$ towers as well as a single ghost tower. 
Their heat kernel becomes
\begin{eqnarray}
\label{eqn:KKbndy}
&&K^{\textrm{KK.bndy}}_4=\frac{1}{2\pi}\frac{1}{4\pi}\left(3\sum^{\infty}_{k=0}(2k+1)e^{-sk(k+1)}+\sum^{\infty}_{k=0}(2k+1)e^{-s(k-1)k}\right.\cr
&&\left.+\sum^{\infty}_{k=1}(2k+1)e^{-s(k+1)(k+2)}+e^{-2s}\sum^{\infty}_{k=1}(2k+1)e^{-sk(k+1)}-e^{-2s}\sum^{\infty}_{k=0}(2k+1)e^{-sk(k+1)}\right)\cr
&&=\frac{1}{8\pi^2}\left(5\sum^{\infty}_{k=0}(2k+1)e^{-sk(k+1)}+2-2e^{-2s}\right)\cr
&&=\frac{1}{8\pi^2}\left(\frac{5}{s}+\frac{5}{3}+\frac{13}{3}s+\mathcal{O}(s^2)\right)~.
\end{eqnarray}
Reading off the constant term $K^{\textrm{KK.bndy}}_4$ we find $a^{\textrm{KK.bndy}}=\frac{5}{12}$ from 
\eqref{eqn:adef}. Adding the bulk contribution $a^{\text{KK.bulk}}=\frac{1}{72}$, we get $a^{\text{KK.bulk+bndy}}=\frac{31}{72}$, which also agrees with the result in \cite{Castro:2018hsc}.

\bigskip
\noindent
{\bf \emph{Gravitino block} }
    
According to table \ref{summary:h:bndy}, the gravitino block comprises $4$ towers of boundary modes with $m_{l}^{2}=(l+1)^{2}-\frac{1}{4}$, $l\geq 0$, each with degenracy $2l+2$. This spectrum gives the heat kernel 
\begin{eqnarray}
K^{\text{gravitino.bndy}}_{4}&=&K^{\text{zero}}_{A}K_{S}\nonumber\\
&=&-\frac{1}{2\pi}~\frac{4}{4\pi}\sum_{k=0}(2k+2)e^{-\left((k+1)^2-\frac{1}{4}\right)s}\nonumber\\
&=&-\frac{4}{8\pi^{2}s}\left(1-\frac{1}{6}s+\mathcal{O}(s^2)\right)e^{\frac{1}{4}s}\nonumber\\
&=&-\frac{4}{8\pi^{2}s}\left(1+\frac{1}{12}s+\mathcal{O}(s^2)\right)~,
\end{eqnarray}
corresponding to $a^{\text{gravitino.bndy}}=-\frac{1}{12}$. With the bulk contribution $a^{\text{gravitino.bulk}}=-\frac{17}{1440}$. Again, the sum $a^{\text{gravitino.bulk+bndy}}=-\frac{137}{1440}$ agrees with that of \cite{Castro:2018hsc}.

\bigskip
For ${\cal N}=8$ SUGRA, there are 1 KK block, 27 vector blocks, and 4 gravitino blocks. In this case the total contribution from the boundary  modes becomes: 
\begin{eqnarray}
&&a^{\text{boundary}}_{\mathcal{N}=8}=27\times\frac{1}{12}+\frac{5}{12}-4\times\frac{1}{12}=\frac{7}{3}~.
\end{eqnarray}

For ${\cal N}=4$ SUGRA with $n_V$ matter multiplets, there are 1 KK block, ($n_V+5$) vector blocks and 2 gravitino blocks, which give the boundary modes contribution $a^{\text{boundary}}_{\mathcal{N}=4}=\frac{n_V+8}{12}$.
%

\subsection{Zero Mode Corrections}
    
Almost all of the modes we encounter are suppressed in the heat kernel \eqref{eqn:K3def}: their eigenvalue is strictly positive. The zero modes are the exceptions: they are constant on the AdS$_2$ like all boundary modes but they are also constant on the $S^2$; so they are zero-modes on the full spacetime AdS$_2\times S^2$. The canonical relation between the heat kernel and the effective action which is implicitly presumed in the formula \eqref{eqn:adef} for the anomaly coefficient $a$ requires damping for large $s$ of an integral over the Feynman parameter $s$ and this assumption fails in the case of zero-modes. 

The correct treatment of zero-modes takes advantage of their relation to symmetries which means their contributions to the path integral are given by integrals over the volume of the appropriate symmetry group, rather than Gaussian integrals over damped modes \cite{Sen:2011ba}. Therefore, the correct contribution to the conformal anomaly $a$ depends on the dimension of the symmetry parameter which is $\Delta= 1, {3\over 2}, 2$ for vectors, gravitini, tensors. The heat kernel \eqref{eqn:K3def} includes all modes with weight $1$ but the correct scaling dimension is $\Delta$ for bosons and $2\Delta$ for fermions. The zero mode {\it correction} takes this effect into account. 

Gauge symmetry generators have $\Delta = 1$ so their zero-modes are, by chance, already accounted for correctly in the na\"{i}ve heat kernel, in the sense that the formula for $a$ \eqref{eqn:adef} can be trusted. Moreover, on the nonBPS branch the gravitino has no zero-modes, because supersymmetry is entirely broken. Therefore,  the KK-block is the only one affected by zero mode corrections. For diffeomorphisms $\Delta=2$ so, since they were already counted with weight one, the contributions of these zero modes should be doubled.
     
In the KK block, there are in total $6$ zero modes from non-normalizable diffeomorphisms that need zero mode corrections: 
$3$ zero modes from the AdS$_2$ tensor $H_{\{\mu\nu\}}$ and $3$ more from the mixed vector modes \eqref{eqn:mixgauge} with $l=1$. This gives the zero mode correction
\begin{eqnarray}
a^{\textrm{KK.zero}}=2\pi^2\times6\times\frac{1}{8\pi^2}=\frac{3}{2}
\end{eqnarray}
This is the same as the contribution from $m^2=0$ modes to the sum \eqref{eqn:KKbndy} over KK boundary modes. Thus, by adding this zero mode {\it correction} their contribution is doubled, as it should be. 

\subsection{Summary of Anomaly Coefficients}
\renewcommand{\arraystretch}{1.5}
\begin{table}[htbp]
\centering
\begin{tabular}{|c|c||c|c|c||c|c|}
\hline
Blocks & d.o.f. &	$a^{\text{bulk}}$  & $a^{\text{bndy}}$ & $a^{\text{zero}}$	& $a^{\text{bulk}}_{\text{+bndy}}$	& $a^{\text{total}}$\\\hline\hline
Scalar & 1&$\frac{1}{360}$ & $0$ & $0$ & $\frac{1}{360}$ & $\frac{1}{360}$\\\hline
Gaugino &	4&$-\frac{17}{2880}$ & $0$ & $0$ &$-\frac{17}{2880}$ &$-\frac{17}{2880}$\\\hline
Vector & 3&$\frac{1}{120}$ & $\frac{1}{12}$ & $0$ &$\frac{11}{120}$ &$\frac{11}{120}$\\\hline
Gravitino &	8&$-\frac{17}{1440}$ & $-\frac{1}{12}$ & $0$ & $-\frac{137}{1440}$ & $-\frac{137}{1440}$\\\hline
KK & 5	&$\frac{1}{72}$ & $\frac{5}{12}$ & $\frac{3}{2}$ & $\frac{31}{72}$ & $\frac{139}{72}$\\\hline\hline
$\mathcal{N}=4$ & $32+16n_V$ & $\frac{n_V+2}{96}$ & $\frac{n_V+8}{12}$ & $\frac{3}{2}$ & $\frac{3n_V+22}{32}$ & $\frac{3n_V+70}{32}$\\
 \hline
 $\mathcal{N}=8$ &256& $\frac{1}{6}$ & $\frac{7}{3}$ & $\frac{3}{2}$ & $\frac{5}{2}$ & $4$\\
 \hline
 \end{tabular}
\caption{Anomaly Coefficients of the Non-BPS Blocks.}
\label{eqn:atable}
\end{table}
\renewcommand{\arraystretch}{1}
As summary of this section we give our results for the anomaly coefficients $a$ in table \ref{eqn:atable}. The entry for boundary modes
$a^{\text{bndy}}$ includes na\"{i}ve zero modes and $a^{\text{zero}}$ denotes the corrections determined by the more careful treatment. The sum $a^{\text{bulk}}_{\text{+bndy}}$ is of interest since it can be compared with results from the local method \cite{Castro:2018hsc}. We find agreement for each of the $5$ type of blocks. This gives great confidence in all our computations. 


\section{Compactifications with an AdS$_3$ Factor}
\label{section:3D to 2D}
In this section we consider the special case where the AdS$_2\times S^2$ geometry arises from AdS$_3\times S^2$ with $(0,4)$ supersymmetry through a reduction along a direction that is nearly null. We recover the black hole spectrum on the BPS (or nonBPS) branch depending on whether the reduction is along the ``0" (or the ``4") direction.

\subsection{String Theory on AdS$_3\times S^2\times {\cal M}$}
We consider M-theory compactified to 5D on a Calabi-Yau manifold $\cal{M}$ in the supergravity limit. The 5D ${\cal N}=2$ content of this theory was worked out in \cite{Cadavid:1995bk}. We include cases with enhanced holonomy ${\cal M}=K3\times T^2$ and ${\cal M}=T^6$ so, in the long distance approximation, we effectively study 5D SUGRA with ${\cal N}\geq 2$ supersymmetry. 
It is useful to describe this theory as ${\cal N}=2$ SUGRA coupled to $n_S={\cal N}-2$ gravitino multiplets (corresponding to supersymmetry extended beyond ${\cal N}=2$) and also to ${\cal N}=2$ matter in $n_V$ vector multiplets and $n_H$ hypermultiplets.

In the setting of these 5D theories we consider field configurations with magnetic fluxes through an $S^2$. They correspond to black string solutions in 5D that are interesting for our purposes because, after further compactification of the string on a circle, they correspond to black holes in 4D \cite{Maldacena:1997de}. We focus on fluxes such that the world-volume of the 5D black strings preserve $(0,4)$ supersymmetry while their gravitational description features an AdS$_3\times S^2$ near horizon geometry. Supergravity fluctuations in this background can be
classified by the quantum numbers of primary fields $(h_L, h_R; j_R)$, where $h_L$ is the scaling dimension with respect to an $SL(2)_L$ isometry of AdS$_3$ and $h_R$, $j_R$ are the quantum number under $SL(2)_R$ and $SU(2)_R$ isometries of AdS$_3$ and $S^2$, respectively.

Because the 5D black string solution preserves $(0,4)$ supersymmetry we can organize its spectrum into supermultiplets. The supergravity fluctuations are all in short multiplets characterized by chiral primaries (states with $h_R=j_R$ but any $h_L$) and their descendants under the preserved ${\cal N}=2$ supersymmetry are
\begin{equation}
\label{eqn:shortmulti}
(h_L, h_R; j_R)~,
\quad
2(h_L , h_R+{1\over 2}; j_R -{1\over 2})~,
\quad (h_L, h_R+1; j_R-1)~,
\end{equation}
with appropriate truncations of the multiplet for small values of $j_R$. The short multiplet numerically has $h_R=j_R$ but we retain both notations to emphasize that these are quantum numbers of two distinct operators. The short multiplet structure applies to all fluctuations in the supergravity approximation so it is common to present the black hole spectrum in terms of the chiral primaries, with descendants under supersymmmetry \eqref{eqn:shortmulti} implied. A standard computation (see e.g. \cite{deBoer:1998kjm}) yields the spectrum of chiral primaries for the $AdS_3\times S^2$ compactification of 5D supergravity given in table \ref{table:5d_hhbarjbar}. We want to deduce the implications of this spectrum on AdS$_3\times S^2$ for theories on AdS$_2\times S^2$.

\renewcommand{\arraystretch}{1.75}
    \begin{table}[h]
	\centering
    \scalebox{0.95}[0.95]{
	\begin{tabular}{|c|c|}
		\hline
		5D multiplets &	Spectrum $(h_L, h_R, j_R)$ of chiral primaries ($h_R=j_R$) \\\hline\hline
		Hyper &	$2(k+1,k+\frac{1}{2}; k+\frac{1}{2})$ \\\hline
		Vector & $(k+2,k+1; k+1) \quad (k+1,k+1; k+1)$ \\\hline
		Gravitino &	$(k+2,k+\frac{1}{2}; k+\frac{1}{2}) \quad (k+2,k+\frac{3}{2}; k+\frac{3}{2}) \quad(k+1,k+\frac{3}{2}; k+\frac{3}{2})$  \\\hline
		Gravity & 	$(k+3,k+1; k+1) \quad (k+2,k+1; k+1) \quad (k+2,k+2; k+2) \quad (k+1,k+2; k+2)$ \\
		\hline
	\end{tabular}}
	\caption{The spectrum of chiral primaries on $AdS_3\times S^2\times \mathcal{M}$. The label $k=0,1,\ldots$}\label{table:5d_hhbarjbar}
\end{table}
\renewcommand{\arraystretch}{1}

\subsection{nNull Reduction: Thermodynamics}
\label{subsec:nnullT}

Many versions of the reduction from AdS$_3$/CFT$_2$ to AdS$_2$/CFT$_1$ have appeared in the literature over the years, including \cite{Gupta:2008ki,Balasubramanian:2009bg,Castro:2010vi,Cvetic:2016eiv,Hartong:2017bwq}. However, the recent advent of nAdS$_2$/nCFT$_1$ correspondence \cite{Almheiri:2014cka,Maldacena:2016upp} justifies renewed scrutiny of this point.

We first describe the dimensional reduction from a thermodynamic point of view, that is more familiar. Because of the chiral nature of CFT$_2$'s it is useful to introduce two independent ``temperatures" $T_{L,R}$ that incorporate both ``the" temperature $T$ (the thermodynamic potential for energy $E = (h_L + h_R) / \ell_3$)
\begin{equation}
{1\over T} = {1\over 2} \left( {1\over T_L}  + {1\over T_R} \right)~,
\end{equation}
and an independent chemical potential (the difference of ``temperatures") for the spin $s = h_L- h_R$.

Implementing the low temperature limit $T\to 0$ by taking $T_R\to 0$ with $T_L$ fixed, the semiclassical entropy of the theory takes the form
\begin{equation}
\label{eqn:nearestS}
S = {\pi^2\over 3} \left( c_L T_L + c_R T_R\right)\ell_3  =  S_0 + {1\over 2}\pi  T {\cal L}  + {\cal O} ( T^2 ) ~,
\end{equation}
where the extremal entropy $S_0= {\pi^2\over 3} c_L T_L\ell_3$ is independent of the temperature and the length scale ${\cal L} = {2\pi\over 3} c_R \ell_3$ that characterizes the linear term in the temperature is proportional to the inverse mass gap of the theory \cite{Preskill:1991tb,Maldacena:2016upp,Almheiri:2016fws}. Our normalization for the length scale ${\cal L}$ follows \cite{Larsen:2018iou} and ensures that it agrees with the ``long string scale" that is characteristic of the $(0,4)$ models underlying microscopics of 4D black holes. 

The {\it strict} extremal limit $T\to 0$ clearly retains states of the form $|{\rm anything}, {\rm gs}\rangle$ where the $R$-sector is in its ground state (except perhaps for a finite ground state multiplicity) and ``anything" is the origin of the extremal entropy $S_0$. In the standard BPS limit ``anything" are the states counted by the elliptic genus.

The {\it near} extremal limit is qualitatively different: it is the theory of {\it excitations above the strict extremal limit} $T\to 0$. If focusses on states that take the schematic form $|{\rm anything}, \delta {\rm gs}\rangle$. The right-moving excitations $|\delta {\rm gs}\rangle$ are responsible for the term in the entropy \eqref{eqn:nearestS} that is linear in $T$. It is the spectrum of these excitations that we study.

The upshot of our discussion of near-extreme thermodynamics is that reduction from AdS$_3\times S^2$ to AdS$_2\times S^2$ amounts to a basic prescription: simply {\it disregard} the left moving weight $h_L$ corresponding to the ``anything" that specifies the extremal state and retain the right moving weight $h_R$ that characterizes the excitation.
Simple as this algorithm may be, it is quite unusual. The canonical set-up for Kaluza-Klein compactification considers a small Kaluza-Klein circle $S^1$ and finds that the low energy approximation retains only modes that are {\it constant} on the compactification circle because higher Fourier modes on the $S^1$ are ``heavy". In contrast, our prescription keeps {\it all} modes on the Kaluza-Klein circle, we {\it omit} a ``momentum" quantum number rather than insisting that it vanishes.

The nNull reduction is chiral in that it (nearly) projects to either the L(eft) or the R(ight) moving sector, depending on whether we study $T_L\to 0$ or $T_R\to 0$. Its two versions are equivalent a priori but, when we apply the construction to the $(0,4)$ CFT$_2$'s that we have in mind, there is an asymmetry between the two chiralities. In this subsection, we elected to focus on the nNull reduction $T_R\to 0$ that (nearly) projects on the {\it BPS branch}, since that facilitates comparison with the literature. However, our interest in this paper will ultimtaely is primarily in the analogous discussion for the nonBPS branch. It follows by interchanging $L$ and $R$ labels.

\subsection{nNull Reduction: Kinematics}
\label{subsec:nnullK}

The thermodynamic reasoning above establishes features that reduction from AdS$_3$/CFT$_2$ to AdS$_2$/CFT$_1$ must exhibit in order to describe the facts we have established by explicit computations in AdS$_2\times S^2$. They are not consistent with standard Kalaza-Klein reduction on a spatial circle so their geometrical implementation must be nonstandard. In the folliowing we show that they can be recovered from {\it null reduction}, {\it i.e.} ``compactification" on a null circle. The details will not only prove illuminating conceptually but also yield precise consequences that we can test. 

A Lorentzian CFT$_2$ on a spatial circle with radius $R$ is obviously invariant under simultaneous shift of the two null coordinates $x_{R,L} = t \pm x$ by $\pm 2\pi R$. However, due to invariance under a boost (with rapidity $\eta$) it is also invariant under shifts of these null coordinates by unequal amounts $\pm 2\pi R e^{\pm\eta}$.
Therefore, as in the DLCQ description of M(atrix)-theory \cite{Seiberg:1997ad,Balasubramanian:2009bg}, there is a family of equivalent theories that all have the same fixed periodicity of the coordinate $x_L$ but variable periodicity of $x_R$. As this periodicity get smaller, states with large ``momentum" $h_R$ become heavy, as the intuition from standard Kaluza-Klein compactification suggests. However, in contrast to the standard construction, the value of the ``momentum" $h_L$ is inconsequential in this limit. 

In the language of effective quantum field theory, the nNull-reduction presents operators in the theory as
\begin{equation}
\left({p_R\over\Lambda}\right)^{h_R-1} {\cal O}^{(h_L,h_R;j_R)}(x_L,x_R)~,
\label{eqn:chiraleft}
\end{equation}
where $p_R$ is the typical frequency corresponding to the $x_R$ dependence and $\Lambda$ is the ``R" cutoff. The dependence on $x_L$ is inconsequential. The strict IR limit takes the cut-off $\Lambda\to\infty$ with the physical momenta $p_{L,R}$ fixed so only operators with $h_R=1$ remain. These ground states of the R sector are the BPS states in the case of a CFT$_2$ with $(0,4)$ supersymmetry. These important operators form the chiral ring of the CFT$_2$ and they are counted by the the elliptic genus. However, the {\it near} IR limit describes the {\it approach to} the IR limit by operators \eqref{eqn:chiraleft} with $h_R>1$. Geometrically, this corresponds to compactification along a direction that is {\it nearly} null. We refer to this construction as a nNull reduction.

In the nNull reduction procedure, the wave functions on AdS$_3$ generally depend on the $x_L$ coordinate but we are instructed to ignore this dependence and instead focus exclusively on the $R$ direction. Therefore, the effective 2D wave functions that follow from nNull reduction depend on the position in AdS$_2$. We interpret our computations directly in 2D as the identification of this dependence.

The nNull reduction thus ignores the $L$ sector and describes the dynamics of the $R$ sector as a self-contained theory. It is a consistency condition on this procedure that operators with identical $x_R$ dependence but distinct $x_L$ dependence realize physics that is largely independent of the latter. This is indeed the expectation: the $L$ sector is in a thermal state characterized by temperature $T_L$ and, according to standard arguments in statistical mechanics, the precise state of this thermal background is inconsequential.

The situation is similar to the well-known description of quasiparticles in the effective field theory of Fermi liquids. In that context the vast majority of the electrons reside deep under the Fermi surface but these ``typical" electrons are not the interesting ones: the nontrivial dynamics is captured by the quasiparticles corresponding to low energy excitations on top of the Fermi surface. It is consistent that the Fermi liquid theory ignores the vast number of states under the Fermi surface as long as the quasiparticles are long lived, a condition that is satisfied at low temperature. Similarly, in our black hole context, the coupling between left- and right-moving sectors will also be suppressed thermally. We can interpret the small residual interaction as the origin of Hawking radiation from the black hole \cite{Das:1996wn}.

\subsection{Explicit Comparison Between AdS$_3\times S^2$ and AdS$_2\times S^2$}
We can use the prescription from the preceding subsection to compare results from explicit computations in 4D with dimensional reductions from 5D. It is important to distinguish two cases from the 4D point of view: the BPS branch that was already discussed in the literature \cite{Michelson:1999kn,Corley:1999uz} and the nonBPS branch that this paper analyzes in detail. They correspond to two distinct dimensional reductions of the spectrum on AdS$_3\times S^2$. In terms of the labels $(h_L, h_R; j_R)$ employed in table \ref{table:5d_hhbarjbar} for bulk 5D representations they are:
\begin{itemize}
\item
The {\it BPS branch}: the dimensional reduction removes the $h_L$ quantum number. It is manifest that the spectrum is organized into short multiplets of the form \eqref{eqn:shortmulti} also after reduction. Starting from the 5D spectrum in table \ref{table:5d_hhbarjbar} we recover the bulk BPS spectrum on AdS$_2\times S^2$ presented in table \ref{summary:h_BPS} for reference and comparison.
\item
The {\it nonBPS branch}:
the reduction removes the $h_R$ quantum number from the labels $(h_L, h_R; j_R)$. Thus, to find the spectrum on the nonBPS branch of AdS$_2\times S^2$ we first augment the chiral primaries in table \ref{table:5d_hhbarjbar} with the structure of short multiplets \eqref{eqn:shortmulti} and only then omit the index $h_R$. The spectrum of primaries that follows from this procedure retains no simplifications that can be obviously traced to supersymmetry. Nonetheless, the result for primaries identified this way agree with our explicit computations on AdS$_2\times S^2$ presented in table \ref{summary:h}.
\end{itemize}

	\renewcommand{\arraystretch}{1.75}
    \begin{table}[h]
	\centering
	\begin{tabular}{|c|c|c|}
		\hline
		4D supermultiplet &	Spectrum $(h,j)$ of BPS solutions	& $SU(6)$ \\\hline\hline
		Hypermultiplet &  $2(k+\frac{1}{2},k+\frac{1}{2}) \qquad 4(k+1,k) \qquad 2(k+\frac{5}{2},k+\frac{1}{2})$ & {\bf 20}  \\\hline
		Vector multiplet & $2(k+1,k+1) \qquad 4(k+\frac{3}{2},k+\frac{1}{2})\qquad 2(k+2,k) $& {\bf 15}  \\\hline
		Gravitino multiplet &	$2(k+\frac{3}{2},k+\frac{3}{2}) \qquad 4(k+2,k+1) \qquad 2(k+\frac{5}{2},k+\frac{1}{2}) $ & {\bf 6}  \\\hline
		Gravity multiplet & 	$2(k+2,k+2) \qquad 4(k+\frac{5}{2},k+\frac{3}{2})\qquad 2(k+3,k+1)$& {\bf 1}  \\
		\hline
	\end{tabular}
	\caption{Bulk spectrum of BPS solutions. The integral label $k\geq0$. In each line the first entry is the chiral primary and the remaining entries reflect the structure \eqref{eqn:shortmulti} of a short multiplet.}
	\label{summary:h_BPS}
\end{table}

In the discussion of CFT$_2$'s in this paper we have assigned the theory $(0,4)$ supersymmetry. This convention implies no loss of generality by itself but, once we have it, it is consequential that in subsections \ref{subsec:nnullT} and \ref{subsec:nnullK} we discussed reduction along the null-direction with label $L$, corresponding to the thermodynamic limit $T_R\to 0$. This choice preserves supersymmetry so it amounts to focus the BPS branch of AdS$_2\times S^2$. The discussion of the nonBPS branch is entirely analogous but, as noted in the end of subsection \ref{subsec:nnullT}, the labels $L$ and $R$ must be interchanged throughout. In the introduction we similarly opted to assign labels $L, R$ such that they are appropriate for the more familiar BPS branch.

With these potential confusions in mind, we spell out the details for each 5D ${\cal N}=2$ multiplet at a time:
\begin{itemize}
\item \emph{Hypermultiplet}\\
The on-shell field content of a 5D hypermultiplet in ${\cal N}=2$ supergravity is two gaugini ($2\times2$ d.o.f), and four scalars ($4 \times 1$ d.o.f.).
On the BPS branch this amounts precisely to a 4D hypermultiplet but on the nonBPS branch the fields split so fermions are in one gaugino block (with two gaugini) and the bosons are in four scalar blocks (each with one real scalar).

Table \ref{table:5d_hhbarjbar} indicates that on AdS$_3\times S^2$ an ${\cal N}=2$ hypermultiplet is organized in two towers of chiral primaries
that both have $(h_L,h_R;j_R)=(k+1,k+{1\over 2}; k+{1\over 2})$ where $k=0, 1, \ldots$. The structure of short multiplets given in \eqref{eqn:shortmulti} then yields $8$ towers of primary fields with $(h_L,h_R;j_R)=2(k+1,k+{1\over 2}; k+{1\over 2}), 4(k+1,k+1; k), 2(k+1,k+{3\over 2}; k-{1\over 2})$. In the last towers the entry with $k=0$ is empty so we may replace these tower with $2(k+2,k+{5\over 2}; k+{1\over 2})$ with $k=0, 1, \ldots$.

Dimensional reduction to the BPS branch of AdS$_2\times S^2$ simply omits $h_L$. The resulting $8$ towers indeed reproduce the BPS spectrum found directly in 4D that is summarized in table \ref{summary:h_BPS} \cite{deBoer:1998kjm,Larsen:1998xm}.

On the nonBPS branch we must instead remove the quantum number $h_R$. This results in $4$ bosonic towers with the quantum numbers given
in table \ref{summary:h} for a scalar block, i.e. a minimally coupled scalar field. Importantly, it also gives $4$ fermion towers with the assignments previously found for a gaugino block on the nonBPS branch.

\item \emph{Vector multiplet}\\
The on-shell field content of a 5D vector multiplet in ${\cal N}=2$ supergravity is one 5D vector field (3 d.o.f.), two gaugini ($2\times2$ d.o.f), and one scalar (1 d.o.f.). Dimensional reduction of a 5D vector field gives a 4D vector field and a {\it real} scalar so an ${\cal N}=2$ vector multiplet in 5D corresponds to an ${\cal N}=2$ vector multiplet in 4D on the BPS branch, comprising one 4D vector, two gaugini and a {\it complex} scalar. On the nonBPS branch these $8$ degrees of freedom are organized into one vector block (a 4D vector plus one real scalar), one gaugino block (two gaugini), and one scalar block (one real scalar).

On AdS$_3\times S^2$ an ${\cal N}=2$ vector multiplet gives chiral primaries that, according to table \ref{table:5d_hhbarjbar}, are organized in two towers with $(h_L,h_R;j_R)=(k+2,k+1; k+1)$ and $(k+1,k+1; k+1)$ where $k=0, 1, \ldots$. The structure of short multiplets given in \eqref{eqn:shortmulti} then yields $8$ towers of primary fields.

On the BPS branch our algorithm instructs us to omit the $h_L$ index so it is immediately clear that the reduction of the 5D spectrum to AdS$_2\times S^2$ yields two copies of $(h_R;j_R)=(k+1; k+1)$, each with the descendants prescribed by \eqref{eqn:shortmulti}. This agrees with the BPS result exhibited in table \ref{summary:h_BPS}.

The nonBPS branch is less familiar, but equally simple. Upon omission of the quantum number $h_R$, the $8$ aforementioned towers of primary fields each give unambiguous values for the pair $(h_L, j_R)$. The quantum numbers found by this procedure can be organized into the sum of the spectra presented in table \ref{summary:h} for a vector block, a gaugino block, and a scalar block.

\item	\emph{Gravitino multiplet}\\
The 5D gravitino multiplet consists of one 5D gravitino ($4$ d.o.f.), two 5D vectors ($2\times3$ d.o.f.) and a gaugino ($2$ d.o.f.). Dimensional reduction of a 5D gravitino gives a gravitino and a gaugino in 4D. An ${\cal N}=2$ gravitino multiplet in 5D therefore corresponds to one 4D gravitino, two 4D vectors, two gaugini, and two scalars. On the BPS branch these fields amount to the sum of an ${\cal N}=2$ gravitino multiplet and an ${\cal N}=2$ ${1\over 2}$-hypermultiplet in 4D. However, on the nonBPS branch, they decompose as the sum of half a gravitino block (one gravitino plus one gaugino in 4D), two vector blocks (two vectors plus two scalars in 4D) and half a gaugino block (one gaugino).

The 5D quantum numbers on AdS$_3\times S^2$ given in table \ref{table:5d_hhbarjbar} indeed reduce to the sum of a gravitino multiplet and half a hypermultiplet entries given for the 4D BPS branch in table \ref{summary:h_BPS}, upon omission of the $h_L$ index. After omission of the $h_R$ index they similarly agree with the sum of half a gravitino block, two vector blocks, and half a gaugino block given for the 4D nonBPS branch in table \ref{summary:h} .

\item	\emph{Gravity multiplet}\\
The gravity multiplet in 5D ${\cal N}=2$ SUGRA consists of the 5D graviton ($5$ d.o.f.), two 5D gravitini ($2\times4$ d.o.f), and the 5D graviphoton ($3$ d.o.f). On the BPS branch these fields are represented in 4D as the sum of an ${\cal N}=2$ gravity multiplet ($4+4$ d.o.f.) and an ${\cal N}=2$ vector multiplet ($4+4$ d.o.f.). On the nonBPS branch, they are represented instead as the sum of a KK-block ($5$ d.o.f.), one gravitino block ($2\times4$ d.o.f), and a 4D vector block ($3$ d.o.f).

The 5D quantum numbers on AdS$_3\times S^2$ given in table \ref{table:5d_hhbarjbar} for the gravity multiplet indeed reduce to the sum of the gravity  and hypermultiplet entries given for the 4D BPS branch in table \ref{summary:h_BPS}, upon omission of the $h_L$ index. After omitting the $h_R$ index they similarly agree with the sum of a KK block, two gravitino blocks, and a vector block given for the 4D nonBPS branch in table \ref{summary:h}.
\end{itemize}

It is interesting that the decomposition into decoupled blocks on the nonBPS branch faithfully reflect their 5D origin: the 5D graviton reduces to the KK block, the two 5D gravitini reduce to a gravitino block, and the 5D vector field reduces to the vector block. 

The dimensional reduction from 5D to 4D illuminates the unsettling feature that fermions on the nonBPS branch all have integral conformal weight in AdS$_2$. A 5D spinor on AdS$_3\times S^2$ has half-integral spin on AdS$_3$ and $S^2$ independently. Projection of the half-integral spin vector in AdS$_3$ on to the periodic spatial coordinate give a half-integral value of $s=h_L - h_R$. Since $h_R$ is tied by supersymmetry to the half-integral spin $j_R$ on $S^2$ it must be that $h_L$ is integral. Since the reduction from AdS$_3$ to AdS$_2$ on the nonBPS branch omits $h_R$ we see that ``the" conformal weight on AdS$_2$ is the integral $h_L$. The integral weights in 2D are therefore perfectly consistent with the spin-statistics theorem.  Indeed, on the nonBPS branch they are required by its 5D version. 

Theories on AdS$_2\times S^2$ that arise through dimensional reduction from AdS$_3\times S^2$ are not the most general ones, specific assumptions on the moduli of the 5D theory must be imposed. However, for the purpose of computing primary fields in supergravity, this situation does not imply any limitations. This is obvious from a practical point of view: there is a canonical equivalence between the allowed supermultiplets of ${\cal N}=2$ supergravity in 4D and in 5D to the extent that, allowing ourselves some abuse of terminology, we apply identical names to analogous representations in 4D and in 5D: supergravity, gravitino, vector, hyper. Therefore, since consistency requires that the black hole spectrum agrees for the  AdS$_2\times S^2$ theories that descend from AdS$_3\times S^2$, it must in fact agree for all black holes. A more abstract approach reaches the same conclusion: since chiral primaries are robust under motions in moduli space it is sufficient to establish the correspondence when AdS$_2\times S^2$ descends from AdS$_3\times S^2$ and then we can conclude that the chiral primaries determined these two ways must agree. From either point of view our explicit computation of the black hole spectrum on the nonBPS branch at some level amounts to a consistency check, albeit a rather nontrivial one.


\section{Global Supersymmetry}
\label{section:global symmetry}
Although our focus is on black holes that do not preserve any supersymmetry it is significant that they are solutions to supergravity. One aspect of this setting is that a remnant of the symmetry persists in the spectrum where it acts as a global supersymmetry.

\subsection{Global Supercharges: the BPS Branch of {\cal N}=8 Theory}

Recall that on the BPS branch there are two spinors $\epsilon_{1,2}$ such that the supersymmetry
transformation \eqref{eqn:deltapsiab} vanishes. This indicates preserved {\it local} supersymmetry and forces the black hole spectrum into short
multiplets with the structure \eqref{eqn:shortmulti}. The nonBPS branch has no analogous symmetries and so its spectrum is not organized into short multiplets. However, on both branches we can exploit the {\it global} part of supersymmetry, i.e. the actions of the transformations \eqref{eqn:deltapsiab} (and analogous actions on the bosons) that do not depend on spacetime position.

On the BPS branch of ${\cal N}=8$ SUGRA the R-symmetry is partially broken as
$SU(8)_R\to SU(2)_R\times SU(6)$. The $2$ preserved and the $6$ broken supersymmetries transform as
$({\bf 2},{\bf 1})$ and $({\bf 1},{\bf 6})$ under the unbroken $SU(2)_R\times SU(6)$. In this section we write the generators of the broken supersymmetry
as $Q^{({1\over 2},\frac{1}{2})}_{A}$ where superscripts refer to $(h_R,j_R)$ and $A$ is an $SU(6)$ index.
These global supersymmetries (anti)commute with the preserved ones so they leave the structure \eqref{eqn:shortmulti} of short multiplets intact.

The chiral primaries are the first entries in each line of table \ref{summary:h_BPS}. Their multiplicities {\bf 20}, {\bf 15}, {\bf 6}, {\bf 1} can be identified with dimensions of $SU(6)$ representations. For example, the towers of hypermultiplets are in the antisymmetric 3-tensor of $SU(6)$ and
their chiral primaries are gaugini with quantum numbers $(h_R,j_R)=(k+{1\over 2},k+{1\over 2})$ that we can write as $\Lambda^{ABC}_{(k+{1\over 2},k+{1\over 2})}$. With this notation the obvious contractions
\begin{eqnarray}
V^{AB}_{(k+1,k+1)}  &=&  Q^{(\frac{1}{2},\frac{1}{2})}_{C} \Lambda^{ABC}_{(k+{1\over 2},k+{1\over 2})} ~, \\
S^{A}_{(k+{3\over 2},k+{3\over 2})}  &=& {1\over 2} Q^{(\frac{1}{2},\frac{1}{2})}_{B}Q^{(\frac{1}{2},\frac{1}{2})}_{C}\Lambda^{ABC}_{(k+{1\over 2},k+{1\over 2})}~, \\
G_{(k+2,k+2)}  & = & {1\over 6}  Q^{(\frac{1}{2},\frac{1}{2})}_{A}Q^{(\frac{1}{2},\frac{1}{2})}_{B}Q^{(\frac{1}{2},\frac{1}{2})}_{C}\Lambda^{ABC}_{(k+{1\over 2},k+{1\over 2})} ~,
\end{eqnarray}
reproduce the remaining chiral primaries in table \ref{summary:h_BPS}. In each case indices indicate $(h_R,j_R)$ so note that, while generally an $SU(2)$ quantum number $j$ can combine with the $j_R=\frac{1}{2}$ of the supercharge and give $j\pm\frac{1}{2}$, for the broken supersymmetry we select just the upper sign. This defines global supersymmetry as an operator in the ring of chiral primary fields.

\subsection{Global Supercharges: the nonBPS Branch of {\cal N}=8 Theory}

We now apply the analogous considerations to the nonBPS branch of $\mathcal{N}=8$ SUGRA. In this case the local supersymmetry is entirely broken but we can exploit the global supersymmetry that remains. Its manifestation is a set of global charges $Q^{(0,\frac{1}{2})}_{A}$ where the index $A$ denotes the fundamental representation of the preserved global $USp(8)$ symmetry and, as usual,  $(h_L,{j}_R)=(0,\frac{1}{2})$ denote the $SL(2)\times SU(2)$ quantum numbers of the AdS$_2\times S^2$ isometries.

We start with the $42$ moduli, the minimally coupled real scalar fields assembled in a $\bf{42}$ of the global $USp(8)$. We denote this antisymmetric four-tensor of $USp(8)$ as $W^{ABCD}_{(k+1,k)}$. Upon action with the global supercharges we find
\begin{equation}
Q^{(0,\frac{1}{2})}_{A}W^{ABCD}_{(k+1,k)} =\Lambda^{BCD}_{(k+1,k+\frac{1}{2})}\oplus \Lambda^{BCD}_{(k+2,k+\frac{1}{2})}~.\\
\label{eqn:gauginospect}
\end{equation}
In this formula, and generally on the nonBPS branch, we refer by definition to an entire tower with indices $k=0,1,\ldots$. In other words, for a given value of $k$ the product of the $SU(2)$ representations $j_R=k$ and $j_R={1\over 2}$ generally allows the values $j_R=k\pm{1\over 2}$. However, in the special case of $k=0$ the option of ``-" is absent so, for the second tower in \eqref{eqn:gauginospect}, we must shift the indices $k\to k+1$. We stress that, on the nonBPS branch, we take towers for both the ``+" and ``$-$" of $j_R=k\pm{1\over 2}$. This is in contrast with the BPS branch where multiplets are shortened so that only the ``$-$" applies for preserved supersymmetries and only the ``+" is active for broken supersymmetries. 
In the context of the global symmetry group $USp(8)$, the contraction of the antisymmetric four-tensor $\bf{42}$ with the supercharge yields an antisymmetric three-tensor $\bf{48}$. Thus the gaugino spectrum \eqref{eqn:gauginospect} agrees with the one we find by explicit computation in section \ref{section:mass spectrum} and summarized in table \ref{summary:h}.

Action with two global supercharges on the minimal scalar fields similarly gives
\begin{equation}
Q^{(0,\frac{1}{2})}_{A}Q^{(0,\frac{1}{2})}_{B}W^{ABCD}_{(k+1,k)}=V^{CD}_{(k+2,k)}\oplus V^{CD}_{(k+2,k+1)}\oplus V^{CD}_{(k+1,k+1)} ~.
\label{eqn:vectorspect}
\end{equation}
Since supercharges anticommute and the fields are antisymmetric in the indices $A,B,\ldots$, the product of the global supersymmetries is effectively symmetric and so corresponds to spin $1$. Generically the product of spin $1$ and spin $k$ gives three towers with spin $k+1$, $k$, and $k-1$. However, for $k=0$ there is obviously just one tower in this product so, according to our convention that the index $k$ has range $k=0,1,\ldots$, we redefined the label $k\to k+1$ in the first two towers of \eqref{eqn:vectorspect}. Since the two $USp(8)$ indices of the fields $V^{CD}$ place the fields in the $\bf{27}$ of $USp(8)$ we recover the spectrum of a vector block reported in table \ref{summary:h}, as claimed.

For three global supercharges we similarly reason that, when acting on an antisymmetric representation, we effectively multiply spin $k$ of the scalar field with spin ${3\over 2}$ of the generators. This gives the decomposition
\begin{eqnarray}
\hspace{-1.0cm}
Q^{(0,\frac{1}{2})}_{A}Q^{(0,\frac{1}{2})}_{B}Q^{(0,\frac{1}{2})}_{C}W^{ABCD}_{(k+1,k)}&=&
S^{D}_{(k+3,k+\frac{1}{2})}\oplus S^{D}_{(k+2,k+\frac{1}{2})}\oplus S^{D}_{(k+2,k+\frac{3}{2})}\oplus S^{D}_{(k+1,k+\frac{3}{2})}~.
\label{eqn:gravitinospect}
\end{eqnarray}
The smallest values are easily checked by hand: the $h_{L}=1$ state in $W^{ABCD}_{(k+1,k)}$ has $j_R=0$ so, after taking the product with spin ${3\over 2}$ of the generators, we find that the $h_{L} =1$ level has just one state and that state has $j_R={3\over 2}$. The only $h_{L}=1$ on the right hand side is the fourth term for $k=0$ and this term indeed has $j_R={3\over 2}$. Similarly, the $h_L=2$ states on the left hand side arise from the spin composition ${3\over 2}  \otimes  1 ={1\over 2}\oplus{3\over 2}\oplus {5\over 2}$, in agreement with the $j_R$ values of the $k=0$ states in the 2nd and 3rd tower and the $k=1$ state in the 4th tower. The result for the spectrum \eqref{eqn:gravitinospect} generated by global supersymmetry agrees with that given in table \ref{summary:h} for half a gravitino block.

Finally, we act with four global supercharges and get
\begin{eqnarray}
Q^{(0,\frac{1}{2})}_{A}Q^{(0,\frac{1}{2})}_{B}Q^{(0,\frac{1}{2})}_{C}Q^{(0,\frac{1}{2})}_{D}W^{ABCD}_{(k+1,k)}&=&
G_{(k+3,k)}\oplus G_{(k+3,k+1)}\oplus G_{(k+2,k+1)}\oplus G_{(k+2,k+2)}\oplus G_{(k+1,k+2)}~.
\nonumber
\end{eqnarray}
We find the structure of the right hand side by multiplication of spin $2$ and spin $k$, and then adjust the indices on states with $h_L=1$ and $h_L=2$ following the model from the preceding paragraph. Our result matches the spectrum of the KK block given in table \ref{summary:h}, as expected.

\subsection{Global Supercharges in AdS$_3$}
We have shown that the black hole spectrum on the BPS branch is generated by global supercharges $Q^{(\frac{1}{2},\frac{1}{2})}_{A}$
while on the nonBPS branch it is organized by $Q^{(0,\frac{1}{2})}_{A}$. It is interesting to inquire whether these charges acting on the AdS$_2$ spectra can descend from AdS$_3$.

The AdS$_3\times S^2$ near horizon geometry of triply self-intersecting strings in 5D ${\cal N}=8$ theory \cite{Kutasov:1998zh,Larsen:1999dh} features a supercharge of the form $Q^{(0, \frac{1}{2};\frac{1}{2})}_{A}$ where $(h_L, h_R; j_R)= (0, \frac{1}{2};\frac{1}{2})$. According to the rules for dimensional reduction introduced in section \ref{subsec:nnullT} omission of $h_L$ yields the BPS branch while omission of $h_R$ gives the nonBPS branch. Therefore, a single AdS$_3$ supercharge gives appropriate supercharges on both branches of the AdS$_2$ theory.
This construction explains the unusual feature that the supercharge on the nonBPS branch has $h=0$. This is possible because the energy $h_R$ is unimportant after the reduction to the nonBPS branch and is closely related to the reason that fermions have integral conformal weights. 

However, the global symmetry encoded in the index $A$ is not entirely clear. The moduli space of AdS$_3\times S^2$ vacua in 5D ${\cal N}=8$ SUGRA is $F_{4(4)}/USp(2)\times USp(6)$ \cite{Kutasov:1998zh} and from this perspective the index $A$ transforms according to the $USp(2)\times USp(6)$ group in the denominator. Upon dimensional reduction to AdS$_2\times S^2$ this global symmetry must be enhanced to $SU(2) \times SU(6)$ (on the BPS branch) or $USp(8)$ (on the nonBPS branch). It is unsurprising that the global symmetry is enhanced upon restriction to one sector or the other but the details have confusing aspects (see \cite{Larsen:1999dh,Cvetic:2014sxa} for discussion).

\subsection{Global Supersymmetry in the ${\cal N}=4$ Theory: the nonBPS Branch}
It is also interesting to determine the global supersymmetry realized by the spectrum of nonBPS black holes in $\mathcal{N}=4$ SUGRA with $n_V$ matter multiplets. The situation is similar to $\mathcal{N}=8$ SUGRA but for $\mathcal{N}=4$ SUGRA the entire spectrum is not unified into a single representation so we encounter several distinct multiplets.

The structure of global symmetries for the nonBPS branch of $\mathcal{N}=4$ SUGRA with $n_V$ matter multiplets was summarized in table \ref{FluctTable}. The black hole breaks the global symmetry group of the theory $SU(4)_R\times SO(n_V)_{\rm matter}$ to $USp(4)\times SO(n_V-1)_{\rm matter}$ so the global supercharges $Q^{(0,\frac{1}{2})}_{A}$ have $USp(4)$ index $A$. 

\begin{itemize}
\item {\it ${\cal N}=4$ superKK Vector Blocks} 
	
There are $n_V-1$ decoupled blocks in the fundamental of the $SO(n_V-1)$ global symmetry. Each superKK vector block has field content of ${\bf 5}$ scalar blocks, ${\bf 4}$ ${1\over 2}$ gaugino blocks,  and ${\bf 1}$ vector block.
Table \ref{summary:h} gives their spectrum as
\begin{eqnarray}
& 5(k+1,k) \cr
& 4(k+2,k+\frac{1}{2}) ~,~~4(k+1,k+\frac{1}{2})  \cr
& (k+2,k)~,~~ (k+2,k+1)~,~~ (k+1,k+1) ~.
\end{eqnarray}
We can fit this spectrum into a supermultiplet generated by global supercharges $Q^{(0,\frac{1}{2})}_{A}$ acting once or twice on a scalar block $W^{AB}$ in the ${\bf 5}$ of $USp(4)$. The spin-$1$ $USp(4)$ singlet $\Omega^{AB}Q^{(0,\frac{1}{2})}_{A}Q^{(0,\frac{1}{2})}_{B}$ acts trivially in this representation. 

\item{{\it The ${\cal N}=4$ SuperKK Gravity Block}}.

This is the minimal theory with a KK solution: ${\cal N}=4$ SUGRA with $n_V=1$ vector multiplets. Our discussion in section \ref{section:N=8} decomposes the ${\cal N}=4$ matter content into fields that decouple in the KK background: $\bf{1}$ KK block, $\bf{4}$ $\frac{1}{2}$ gravitino blocks, $\bf{6}$ vector blocks, $\bf{4}$ $\frac{1}{2}$ gaugino blocks, and $\bf{1}$ scalar block. Boldfaced letters refers not only to the multiplicity but also to the $USp(4)$ representation. These fields are all singlets of $SO(n_V-1)$ so there is just one ${\cal N}=4$ superKK-block, as expected because gravity is unique. Table \ref{summary:h} gives their spectrum as
\begin{eqnarray}
& (k+1,k) \cr
& 4(k+2,k+\frac{1}{2}) ~,~~4(k+1,k+\frac{1}{2})  \cr
& 6(k+2,k)~,~~ 6(k+2,k+1)~,~~  6(k+1,k+1) \cr
& 4(k+3,k+\frac{1}{2})  ~,~~ 4(k+2,k+\frac{1}{2})  ~,~~ 4(k+2,k+\frac{3}{2})  ~,~~  4(k+1,k+\frac{3}{2}) \cr
& (k+3,k) ~,~~  (k+3,k+1) ~,~~(k+2,k+1) ~,~~ (k+2,k+2)~,~~ (k+1,k+2)~.\nonumber
\end{eqnarray}
We can fit all these fields into a tower of supermultiplets generated by supercharges $Q^{(0,\frac{1}{2})}_{A}$. 
Antisymmetric representations formed by tensoring $0,1,2,3,4$ vectors under the global $USp(4)$ (labelled by  $0,1,2,3,4$  indices $A, B, \ldots$)
account for the degeneracies ${\bf 1}, {\bf 4}, {\bf 6}, {\bf 4}, {\bf 1}$. The middle entry is reducible as an $USp(4)$ representation ${\bf 6}={\bf 5}\oplus {\bf 1}$. However, both components are kept when the singlet $\Omega^{AB}Q^{(0,\frac{1}{2})}_{A}Q^{(0,\frac{1}{2})}_{B}$ is represented 
nontrivially. Moreover, symmetric combinations of $0,1,2,3,4$ supercharges of this form transform as spin $0, {1\over 2}, 1, {3\over 2}, 2$. These spins act on the first line of the equation using the standard product rule of angular momenta and, after compensating for missing entries with small spin by adjusting the index $k$ so $k=0,1,\ldots$ in all cases, the remaining lines follow precisely. 

\end{itemize}

\section*{Acknowledgements}
We thank Alejandra Castro and Victor Godet for useful discussions. This work was supported in part by the U.S. Department of Energy under grant DE-FG02-95ER40899.

\bibliographystyle{JHEP-2}
\providecommand{\href}[2]{#2}\begingroup\raggedright\endgroup

\end{document}